\documentclass[twocolumn, twocolappendix]{aastex631} % 
\usepackage{amsmath, soul}

\newcommand{\um}{$\mu$m}

\newcommand{\rpm}{\raisebox{.2ex}{$\scriptstyle\pm$}}
\newcommand{\vsigma}{v$_{\rm c}$/$\sigma$}

\def\msol{\ifmmode{{\rm M}_{\odot}}\else{M$_{\odot}$}\fi}
\def\mstar{\ifmmode{{\rm M}_{\star}}\else{M$_{\star}$}\fi}
\def\vsigma{\ifmmode{{\rm v}_{\rm c}/\sigma}\else{v$_{\rm c}$/$\sigma$}\fi}
\newcommand{\lir}{L$_{\rm IR}$}

\newcommand{\lsun}{{\rm L$_{\odot}$}}

\newcommand{\td}{T$_{\rm d}$}
\newcommand{\lp}{$\lambda_{\rm peak}$}
\newcommand{\lirlp}{L$_{\rm IR}$--$\lambda_{\rm peak}$}

\newcommand{\iras}{\textit{IRAS}}
\newcommand{\hersch}{\textit{Herschel}}
\newcommand{\spitz}{\textit{Spitzer}}
\newcommand{\wise}{\textit{WISE}}
\newcommand{\hatl}{H-ATLAS}
\newcommand{\sfrsurf}{$\Sigma_{\rm SFR}$}

\newcommand{\lt}{$\rm L_{\rm t}$}

\def\ltsima{$\buildrel < \over \sim$}
\def\simlt{\lower.5ex\hbox{\ltsima}}
\def\gtsima{$\buildrel > \over \sim$}
\def\simgt{\lower.5ex\hbox{\gtsima}}

\newcolumntype{P}[1]{>{\centering\arraybackslash}p{#1}}
\newcolumntype{M}[1]{>{\centering\arraybackslash}m{#1}}

\shorttitle{No Redshift Evolution of Dust Temperatures from $0 < z < 2$}
\shortauthors{P. M. Drew and C. M. Casey}
\begin{document}

\title{No Redshift Evolution of Galaxies' Dust Temperatures Seen from $0 < z < 2$}

\author[0000-0003-3627-7485]{Patrick M. Drew}
\affiliation{Department of Astronomy, The University of Texas at Austin \\ 2515 Speedway Blvd Stop C1400 \\ Austin, TX 78712}

\author[0000-0002-0930-6466]{Caitlin M. Casey}
\affiliation{Department of Astronomy, The University of Texas at Austin \\ 2515 Speedway Blvd Stop C1400 \\ Austin, TX 78712}



\begin{abstract}

Some recent literature has claimed there to be an evolution in galaxies' dust temperatures towards warmer (or colder) spectral energy distributions (SEDs) between low and high redshift. These conclusions are driven by both theoretical models and empirical measurement. Such claims sometimes contradict one another and are prone to biases in samples or SED fitting techniques. What has made direct comparisons difficult is that there is no uniform approach to fitting galaxies' infrared/millimeter SEDs. Here we aim to standardize the measurement of galaxies' dust temperatures with a python-based SED fitting procedure, MCIRSED\footnote{Publicly available at \href{https://github.com/pdrew32/mcirsed}{github.com/pdrew32/mcirsed}}. 
We draw on reference datasets observed by \textit{IRAS}, \textit{Herschel}, and \textsc{Scuba-2} to test for redshift evolution out to $z\sim2$.  We anchor our work to the \lirlp\ plane, where there is an empirically observed anti-correlation between IR luminosity and rest-frame peak wavelength (an observational proxy for luminosity-weighted dust temperature) such that $\left<\lambda_{\rm peak}\right>=\lambda_{\rm t}({\rm L_{\rm IR}}/{\rm L_{\rm t}})^{\eta}$ where $\eta=-0.09\rpm0.01$, L$_{\rm t}=10^{12}$\,\lsun, and $\lambda_{\rm t}=92\rpm2$\,\um.  We find no evidence for redshift evolution of galaxies' temperatures, or \lp, at fixed \lir\ from $0<z<2$ with $>$99.99\% confidence. Our finding does not preclude evolution in dust temperatures at fixed stellar mass, which is expected from a non-evolving \lirlp\ relation and a strongly evolving SFR--M$_\star$ relation. The breadth of dust temperatures ($\sigma_{\log\lambda_{\rm peak}}$) at a given \lir\ is likely driven by variation in galaxies' dust geometries and sizes and does not evolve. Testing for \lirlp\ evolution toward higher redshift ($z\sim5-6$) requires better sampling of galaxies' dust SEDs near their peaks (observed $\sim$200--600\,\um) with $\simlt$1\,mJy sensitivity. This poses a significant challenge to current instrumentation.

\end{abstract}



\section{Introduction}\label{sec:introduction}
Astrophysical dust is a key component of the interstellar medium in
galaxies, despite comprising a negligible proportion of their total
mass \citep[$\simlt$1\%\ of the baryonic budget,
  e.g.][]{remy-ruyer14u}.  Dust is an efficient absorber of
ultraviolet and visible light emitted from both young stars and active
galactic nuclei (AGN). It re-radiates the absorbed energy
thermally between infrared (IR) and millimeter wavelengths
\citep[e.g.][]{Wolfire95q}. The total energy of re-radiated emission,
\lir\ (formally defined as the integral of the SED between
8--1000\,\um), is often used as a fundamental proxy for a galaxy's
star formation rate \citep[e.g.][]{Sanders96a, Kennicutt98a}. Thus, lack of knowledge of a galaxies'
re-radiated dust emission can result in catastrophic
misinterpretation of galaxies' underlying physics.

While rough estimates of a galaxies' infrared luminosity can be made
even with fairly crude IR/mm constraints \citep[e.g.][]{Sanders03c}, the accurate measurement of
galaxies' dust SEDs can unlock a more detailed understanding of the
systems' physical configuration, chemical enrichment, and mass
build-up (see reviews by \citeauthor*{Casey14a} \citeyear{Casey14a}, \citeauthor*{Galliano18o} \citeyear{Galliano18o}, and \citeauthor*{Hodge20a} \citeyear{Hodge20a}).  Beyond IR luminosity, the temperature of thermal dust emission can be roughly inferred using Wien's Displacement Law \citep{Wien96d}, whereby the
rest-frame peak wavelength, \lp, of the SED in $S_{\rm \nu}$ is
inversely proportional to a blackbody's temperature.  Other
characteristics may also be measured -- the optical depth of dust as a function of
wavelength, the dust mass surface density, the emissivity spectral
index $\beta$, the presence of warm dust heated by nuclear processes, the dust extinction curve, and the abundance of Polycyclic Aromatic Hydrocarbons \citep*[PAHs;][]{Galliano18o}. However,
a dearth of photometric constraints in the IR/mm regime for the
majority of galaxies severely limits tight constraints on most of
these quantities, save \lir\ and \lp. 
Additionally, the degeneracies between some physical quantities such as dust emissivities and temperatures further limits our ability to constrain them \citep[e.g.][]{Spilker16a}. Observations that would break these degeneracies do not exist for the majority of galaxies. In the face of this, the most reasonable choice for deciphering trends in the underlying population is to focus on IR luminosities and the wavelengths where the dust emission peaks, which is a proxy for dust temperature; the two parameters encapsulating dust SEDs for which large samples of galaxies have observations.

Measurements of galaxies' dust temperatures in the literature are
plentiful, yet different studies employ different methods for converting a set of
IR/mm photometry to a temperature, ranging from single modified
blackbody fits \citep{Dunne01n}, to a superposition of blackbody fits
\citep{Younger07c, Dale01a, Casey12a}, the use of empirical templates
\citep{Chary01a, Draine07a, Rieke09h} or theoretical models
\citep{Silva98a, Siebenmorgen07a, Burgarella05r, da-cunha15a}.  This
conversion naturally neglects the complexity of galaxies' ISM, where
dust cannot simply be characterized as having one temperature. In
reality, ISM dust exists in several phases, from warm direct
photon-heated dust in close proximity to young OB associations or AGN
to cold, diffuse ISM dust heated only by the ambient radiation field.
In practice, what {\it is} measured from galaxies' SEDs is the
luminosity-weighted dust temperature averaged over the entirety of the
galaxy. Though this reduces the complexity of galaxies' ISM to a single parameter, that parameter is, in principle, directly measurable for a large sample of galaxies spanning cosmic time.

In the simple case
of a modified blackbody, the same set of photometry could be described
as having very different dust temperatures depending on the underlying
dust opacity model. For example, \citet{Cortzen20j} find that an
optically thin blackbody fit to GN20 gives a dust temperature of 33\,K
while an optically thick model gives 52\,K.  For this reason, some
works \citep{Casey18b, Casey18a} have advocated for use of \lp\ to
quantify the dust temperature. \lp\ is a direct, observable quantity
that is model independent and thus a direct reflection of data
constraints. Note that converting \lp\ to a dust temperature still requires assumptions about the dust opacity in the absence of direct constraints.

Given the nuances in both deriving and comparing dust temperatures, it
is not necessarily surprising that the literature has reported several
seemingly contradictory temperature trends with redshift.  The first
set of claims argue that higher redshift galaxies display hotter dust
temperatures on average, in comparison to galaxies at $z\sim0$.
Such conclusions are reached using both observational data at
$z\sim1$--2 \citep[e.g.][]{Magnelli14a, Bethermin15a, Schreiber18a} out to $z\sim4$--5
\citep[e.g.][see also \citeauthor{Bakx21h} \citeyear{Bakx21h} at $z=7.13$]{Magdis12a, Faisst17b, Faisst20a} and via theoretical
modeling, largely at the earliest epochs, $z\,\,\simgt\,\,5$
\citep[e.g.][]{De-Rossi18w, Ma19a, Liang19a, Arata19a, Sommovigo20a}.
In theory, the higher dust temperatures at earlier epochs are
attributed to either a lower relative dust abundance, higher specific
star formation rates (sSFR), or higher star formation surface
densities (\sfrsurf) in high redshift galaxies.  These conditions may
only predominate at $z\,\, \simgt\,\,5$, at which point the dust reservoirs of
even massive galaxies may not have had sufficient time after the Big
Bang to reach an enriched chemical equilibrium.  Observationally,
higher dust temperatures for high redshift galaxies have been measured
in ``main sequence'' galaxies \citep[e.g.][]{Magnelli14a, Bethermin15a, Schreiber18a}
while galaxies elevated above the main sequence are observed to be uniformly `hot.'
Observational work at $z>3$ has focused on galaxies
lying far off expectation \citep{Capak15a} in the relationship between their rest-frame
UV color and the \lir/L$_{\rm UV}$ ratio in the ``IRX--$\beta$'' relationship \citep{Meurer99m} potentially indicative of
different dust compositions. Invoking hotter dust temperatures (hotter
than is often assumed for lower-redshift sources) may bring
observation back in line with expectation \citep[e.g.][]{Bouwens16a, Faisst17b}.

The second set of claims posit that galaxies' dust SEDs evolve toward
{\it colder} temperatures at higher $z$ \citep[e.g.][]{Chapman02g,
  Pope06a, Symeonidis09y, Symeonidis13a, Hwang10a, Kirkpatrick12a,
  Kirkpatrick17a, Magnelli14a}.  They find this to be caused by
higher dust masses \citep[e.g.][]{Kirkpatrick17a}, higher dust
opacities or emissivities, or more extended spatial distributions of
dust at higher redshift \citep[e.g.][]{Symeonidis09y, Elbaz11a,
  Rujopakarn13o}.
While luminous and ultraluminous IR galaxies (LIRGs and ULIRGs) in the
local Universe are compact \citep[$<$1\,kpc; e.g.][]{Condon91k}, higher redshift galaxies with similar
IR luminosities have extended regions of dust emission \citep[$\sim$few\,kpc; e.g.][]{Elbaz11a, Rujopakarn11y, Rujopakarn13o}.  The conjecture of these works is that the
extended distribution of dust in high-$z$ IR luminous galaxies leads to overall cooler dust temperatures than are seen in local (U)LIRGs.

The third set of claims suggest that galaxy SEDs do not, in fact,
evolve substantially with redshift but that any perception in
evolution is due to telescope selection effects.  Such selection
effects have been known in the literature for some time; for example,
850\,\um\ detection may preferentially select the coldest IR luminous
galaxies at any given redshift given the steep temperature dependence of
$S_{\rm 850}\propto T_{d}^{-3.5}$ \citep[e.g.][]{Eales00v, Chapman04a}.
In analyzing several different samples from $0<z<5$, \citet{Casey18a}
find no evidence for evolution in the distribution of galaxies in \lirlp\ space, though this is only argued using aggregated literature fits that do not account for sample incompleteness or biases.  Similarly,
\citet{Dudzeviciute20a} finds no evolution in dust temperture at fixed
IR luminosity for massive dusty galaxies between $1.5 < z < 4$, though their sample is limited to galaxies selected at 850\,\um, and selecting samples at 850\,\um\ will preferentially select for cold sources \citep[e.g.][]{Eales00v, Chapman04b, Casey09q}. Recent works by \citet{Simpson19a} and \citet{Reuter20a} also find no conclusive evidence of evolution.

In this paper we aim to quantitatively address possible evolution in
the \lir--\lp\ relationship from $z\sim0$ to $z\sim2$ and to test the hypothesis that \lir--\lp\ does not evolve over this redshift range. Our goal is to do so using an approach which investigates sample and observing band biases in dust SED fitting in a uniform manner.  To do that we
fit dust SEDs for $\sim$4700 galaxies from three different samples
from the literature to explore telescope selection biases that
may influence measurement of galaxies' dust SEDs and then we place our
best-fit SEDs in context against different claims in the literature.
This paper is organized as follows. In $\S$\ref{sec:sample_selection}
we describe each of the samples we curated for our analysis.  In
$\S$\ref{sec:sed_model} we describe the SED modeling technique we
employ.  In $\S$\ref{sec:lirtd correlation} we describe the
\lirlp\ relation measured first for a $z\sim0$ sample and then go
on to test its redshift evolution using subsequent higher redshift
samples.  Section \ref{sec:discussion} presents a discussion of our
results in context of literature studies, and we make
recommendations for standardizing IR/mm SED fitting procedures for
fair comparison across works.  Finally, we present the MCIRSED tool we
built to carry out our analysis in the appendix ($\S$\ref{sec:bayesian_modeling}).
Throughout this work we adopt the \textit{Planck} 2018 flat cosmology
with $H_0=67.4$\,km\,s$^{-1}$\,Mpc$^{-1}$, and $\Omega_{m} = 0.315$
\citep{Planck-Collaboration20a} and a Kroupa stellar initial mass
function \citep{Kroupa01a, Kroupa03a}.

\section{Data Selection}
\label{sec:sample_selection}

\begin{table*}[]
    \centering
    \caption{Table listing, for each sample, the wavelengths of observation, detection limits, sample size, survey area, redshift range, median redshift, and parent sample references. Uncertainties on the median redshifts are the standard deviations of 1000 bootstrapped calculations of the median redshift.}
    \begin{tabular}{p{0.35\columnwidth}p{0.5\columnwidth}p{0.5\columnwidth}p{0.5\columnwidth}}
        \hline\hline
        Sample Name & \iras\ & \hatl\ & COSMOS \\
        \hline 
        Wavelengths ($\mu$m) & 12 (\wise), 12 (\iras), 22, 25, 60, 100 (\iras), 100 (\hersch), 160, 250, 350, 500 & 22, 100, 160, 250, 350, 500 & 24, 70, 100, 160, 250, 350, 500, 850, 1100, 1200 \vspace{2mm} \\
        Selection of Sample & SNR $>5$ ($>$179\,mJy) at 60\,\um\ and SNR $>4$ ($>$29\,mJy) at 250\,\um\ & SNR $> 3$ ($>$132\,mJy) at 100\,\um\ and SNR $> 4$ ($>$29\,mJy) at 250\,\um & SNR $>3$ in at least two bands: 100\,\um\ ($>$5.34\,mJy), 160\,\um\ ($>$11.6\,mJy), 250\,\um\ ($>$17.4\,mJy), 350\,\um\ ($>$18.9\,mJy), or 500\,\um\ ($>$20.4\,mJy) \vspace{2mm} \\
        Number of Detections (Upper Limits) in Each Band\footnote{The number of galaxies with detections in each band with SNR $\geq$ 3. The number of galaxies with upper limits in each band (SNR $<$ 3) are listed in parentheses as well as the percentage of the total that have upper limits.} & 12\,\um\ (\wise): 502 (9, 2\%), 12\,\um\ (\iras): 16 (495, 97\%), 22\,\um: 378 (133, 26\%), 25\,\um: 65 (446, 87\%), 60\,\um: 511 (0, 0\%), 100\,\um\ (\iras): 364 (147, 29\%), 100\,\um\ (\hersch): 461, (50, 9\%), 160\,\um: 463 (48, 9\%), 250\,\um: 511 (0, 0\%), 350\,\um: 499 (12, 2\%), 500\,\um: 401 (110, 22\%) & 22\,\um: 1656 (511, 24\%), 100\,\um: 2167 (0, 0\%), 160\,\um: 1459 (705, 33\%), 250\,\um: 2167 (0, 0\%), 350\,\um: 1801 (366, 17\%), 500\,\um: 563 (1604, 74\%) & 24\,\um: 1990 (0, 0\%), 70\,\um: 1588 (402, 20\%), 100\,\um: 1666 (324, 12\%), 160\,\um: 1713 (277, 14\%), 250\,\um: 1971 (19, 1\%), 350\,\um: 1827 (163, 9\%), 500\,\um: 1406 (584, 29\%), 850\,\um: 185 (1805, 90\%), 1100\,\um: 18 (1972, 99\%), 1200\,\um: 5 (1985, 99.7\%) \vspace{2mm} \\
        Sample Size (galaxies) & 511 & 2167 & 1990 \vspace{2mm} \\
        Survey Solid Angle (deg$^2$) & 660 & 162 & 2 \vspace{2mm} \\
        Redshift & $0.025 < z < 0.19$ & $0.05 < z < 0.4$ & $0.15 < z < 2.0$ \vspace{2mm} \\
        Median Redshift  & 0.050 $\rpm$ 0.002 & 0.127 $\rpm$ 0.002 & 0.701 $\rpm$ 0.008 \vspace{2mm} \\
        References & \citet{Wang14a, Valiante16a, Bourne16a, Smith17a, Maddox18a, Furlanetto18a} & \citet{Valiante16a, Bourne16a} & \citet{Lee13a}; \citet{Simpson19a}; \citet{Weaver21j} \\
        \hline\hline
    \end{tabular}
    \label{tab:selections}
\end{table*}

\begin{figure}
    \centering
    \includegraphics[width=0.98\columnwidth]{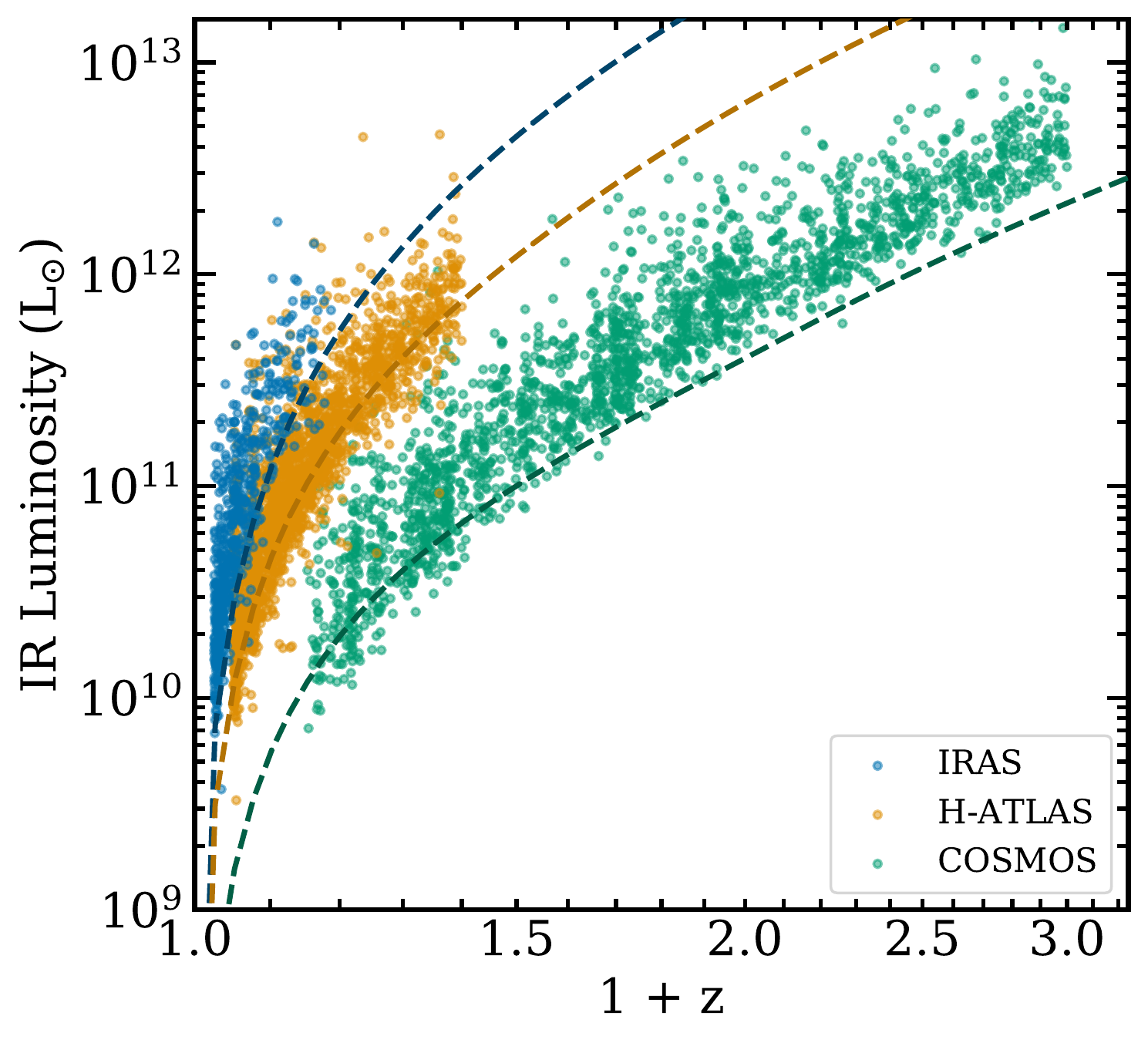}
    \caption{The IR luminosities of galaxies in each sample versus redshift. The \iras\ sample covers the redshift range $0.025 < z < 0.19$, the \hatl\ sample covers $0.05 < z < 0.4$, and the COSMOS sample covers $0.15 < z < 2.0$. The dashed lines correspond to the IR luminosity detectability limits of galaxies at the median peak wavelength (dust temperatures) in each sample. We plot the 60\,\um\ detection limit for the \iras\ sample, the 100\,\um\ detection limit for the \hatl\ sample, and the 160\,\um\ detection limit for the COSMOS sample as they are the most illustrative of the intrinsic selection for each sample. See \autoref{tab:selections} for the total selection function of each sample.}
    \label{fig:z_distribs}
\end{figure}

The aim of this study is to investigate possible redshift evolution in the average dust temperature of galaxies. 
To this end we make use of three different catalogs from the literature that collectively span the redshift range of $0 < z < 2$ giving us a broad dynamic range of IR luminous galaxies that allows us to assemble an unbiased sample with well sampled dust SEDs across a large redshift range.
Each sample has multiple photometric observations at wavelengths spanning roughly the nominal range of thermal dust emission, 8--1000\,\um\ with between six and eleven photometric constraints (with a median of eight constraints).
We require that our samples have observations between rest-frame $40 < \lambda_{\rm peak} < 200$\,\um\ in order to constrain \lp.
This corresponds to a temperature range of 10--80\,K assuming the case of an optically thin blackbody.
We also require that each galaxy have a spectroscopic or high quality photometric redshift derived from optical and infrared (OIR) photometry ($\Delta z/(1+z) \lesssim 0.1$). It is crucial that redshift constraints be independent of the IR/mm photometry, otherwise our results may shield an inherent bias.

Our analysis focuses on three key samples. The first, called the \iras\ sample, is the low redshift sample that forms the anchor understanding of dust SEDs in the local universe. 
The next sample is \hatl\ which extends to $z\sim0.4$, and finally the COSMOS sample which extends to $z\sim2$.
The distribution of each of these samples in \lir--$z$ space is shown in \autoref{fig:z_distribs}, and \autoref{tab:selections} summarizes their defining attributes, further described below.

We limit our analysis to $z < 2$ because galaxies at higher redshifts have significantly lower quality photometric constraints on their dust SEDs. This is primarily due to the lack of sensitivity of \hersch\ SPIRE to ${\rm L_{\rm IR}} \lesssim 10^{13}$\,\lsun\ SEDs at $z > 2$ \citep[e.g.][]{Casey12b}. In addition to low signal-to-noise measurements of $z > 2$ FIR data, $z > 2$ DSFGs tend to have a lower proportion of reliable redshift constraints with a much higher occurance rate of sources lacking OIR counterparts entirely \citep[e.g.][]{Wardlow11v}. In contrast, $\gtrsim$98\% of $z < 2$ DSFGs are detected in 24\,\um\ data and corresponding deep OIR imaging in the COSMOS field \citep[e.g.][]{Magdis12a, Weaver21j} which makes this analysis possible at $z < 2$.

\subsection{IRAS Sample}
\label{subsec:iras_sample}

We build our lowest redshift sample, which we will refer to as the \iras\ sample, by performing a spatial cross match between two separate catalogs.
The first catalog contains Infrared Astronomical Satellite \citep[\iras;][]{Neugebauer84a} sources from the Revised \iras\ Faint Source Redshift Catalog of galaxies \citep[RIFSCz;][]{Wang14a}. This sample is selected at 60\,\um\ and contains observations at 12, 25, 60, and 100\,\um\ covering galactic latitudes $|b| > 20^{\circ}$ in unconfused regions of the sky at 60\,\um\ (where unconfused corresponds to fewer than 8 sources per beam with 6\,\arcmin\ radius) with order of 60,000 sources.
The second catalog is data releases 1 and 2 of the \textit{Herschel} Astrophysical Terahertz Large Area Survey catalog which covers $\sim$120,000 sources \citep[H-ATLAS;][]{Valiante16a, Bourne16a, Smith17a, Maddox18a, Furlanetto18a}. This sample is selected at 250\,\um\ and contains observations at 100, 160, 250, 350, and 500\,\um\ from the SPIRE and PACS instruments covering 660\,deg$^2$.

To create the RIFSCz, \citet{Wang14a} select galaxies detected with \iras\ from the \textit{IRAS} Faint Source Catalog \citep[FSC;][]{Moshir93a} that have ${\rm SNR} > 5$ in the 60\,\um\ band.
The FSC contains $\sim$1.7$\times10^5$ sources with only about 40\% of sources detected at 60\,\um.
In order to obtain accurate positional estimates, \citet{Wang14a} use the likelihood ratio technique \citep[e.g.][]{Sutherland92a} to match FSC sources to 3.4\,\um\ counterparts in the Wide-field Infrared Survey Explorer All-Sky Data Release \citep[\textit{WISE};][]{Wright10a}. This wavelength is used because it is the deepest band in \wise\ and it samples a galaxy's SED at a wavelength where it is likely to be detected from either warm dust or stellar emission mechanisms. The authors then cross match the \iras\ sources with \wise\ astrometry (the later is more precise) and then to several other wide field surveys: the Sloan Digital Sky Survey Data Release 10 \citep[SDSS DR10;][]{Ahn14a} using a matching radius of 3\arcsec, the \textit{Galaxy Evolution Explorer} (\textit{GALEX}) All-Sky Survey Source Catalog \citep[GASC;][]{Seibert12c} using a matching radius of 5\arcsec, the Two Micron All Sky Survey \citep[2MASS][]{Huchra12a} which is already included with the \wise\ catalog, and the first public release of the \textit{Planck} catalog of compact sources \citep[PCCS;][]{Planck-Collaboration13a, Planck-Collaboration14a} using a matching radius of 3\arcmin. Each matching radius used by \citeauthor{Wang14a} is chosen based on the spatial resolution of each match. 
Sources are retained in the RIFSCz catalog even if they are not matched to detections in these ancillary catalogs.
See \citet{Wang14a} for more details.
In the present work focusing on IR SEDs we use \iras\ and \wise\ flux densities  and redshifts for our analysis.

We further cross match RIFSCz sources with the H-ATLAS catalog \citep{Valiante16a, Bourne16a, Smith17a, Maddox18a, Furlanetto18a} to fill out the SEDs at longer wavelengths. \hatl\ presents sources detected at ${\rm SNR} > 4$ at 250\,\um\ (corresponding to a 4$\sigma$ depth of 29\,mJy) in an area of the sky spanning 660\,$\deg^2$.
The catalog contains two data releases: DR1, covering the North Galactic Pole (NGP, 500\,deg$^2$), cross matched with SDSS data release 10 \citep{Ahn14a} for redshifts and optical counterparts and positions, and DR2, covering the South Galactic Pole (SGP, 162\,deg$^2$).

Sources in the RISFCz catalog either have astrometric positions based on the best available precision where reliable counterparts are identified in \citet{Wang14a}: from SDSS DR10, 2MASS, or \wise. From the \citet{Wang14a} astrometry we further cross match RIFSCz sources with both DR1 and DR2 of the \hatl\ catalog using search radii that correspond to radius of the average point spread function of the best available ancillary catalog provided by \citet{Wang14a}. \hatl\ DR1 and the NGP field of DR2 are matched to RIFSCz sources with a search radius of 1\arcsec\ because the RIFSCz catalog in this region is based on SDSS DR10 astrometry, precise to 1\arcsec. However, there is no SDSS DR10 coverage of the SGP field in H-ATLAS DR2, so sources in the SGP field are matched using a less restrictive search radius of 1\farcs6 for counterparts in the 2MASS catalog whose astrometry is slightly less well constrained (154 sources are matched). In the absence of 2MASS counterpart matches, sources from the SGP are matched to the \wise\ astrometry given in the RIFSCz using a matching radius of 3\arcsec (14 sources are matched this way). Note that contamination from false positives are negligible within this search radius given the low source density of both RIFSCz sources ($\sim$1\,deg$^{-2}$) and possible 2MASS counterparts ($\sim$50\,deg$^{-2}$).
Cross matching sources from the RIFSCz to H-ATLAS data releases 1 and 2 results in 780 matches or 2\% of RIFSCz sources, which is the expected proportion of the full RIFSCz catalog because H-ATLAS only covers 2\% of the solid angle spanned by the RIFSCz.

We limit the \iras\ sample to only include galaxies with spectroscopic redshifts (spec-$z$s). This requirement reduces the sample from 780 to 750 galaxies (96\% of the original) spanning the redshift range $z < 0.48$ with median $\left< z\right> = 0.050\rpm 0.002$. This reduction does not show a bias in redshift within the sample.
Requiring spec-$z$s reduces the sample size by only 4\% and thus is a reasonable choice because uncertainties in adopted redshift will translate into uncertainties on our measured rest frame peak wavelengths and IR luminosities.

Because the aim of this work is to look for redshift evolution in the correlation between the peak wavelength and the IR luminosities of galaxies we want a clean snapshot of the \lirlp\ correlation without influence from any potential redshift evolution.
We therefore further restrict the sample to only include galaxies in a redshift range which does not evolve strongly with cosmological time and yet is beyond the local volume.
We limit our sample to galaxies between $0.025 < z < 0.19$.
The lower redshift bound is specifically chosen to remove spatially resolved sources. It corresponds to the redshift where a galaxy with a radius of 10\,kpc would be resolved across two 250\,\um\ \hersch\ beams ($18\arcsec \times 2 = 36\arcsec$).
This removes some of the uncertainty associated with the more extreme aperture corrections required to derive accurate photometric constraints in the local volume.
The upper redshift bound is chosen based on the timeframe over which the star formation rate of galaxies on the star formation ``main sequence'' evolves by less than a factor of two from our lower redshift bound \citep{Speagle14a}.
In other words, the galaxies' average SFRs (at fixed stellar mass) do not evolve by more than two times over this interval. 
This factor of two is chosen because it corresponds to the intrinsic width of the main sequence itself ($\sim0.3$\,dex). Thus, in this snapshot in time the main sequence does not evolve by more than its intrinsic width.
This additional redshift restriction leaves 511 out of 750 galaxies remaining in the sample (68\%).
We also tested setting the lower redshift bound to the redshift where a galaxy with radius 10\,kpc fits within a single beam (at $z=0.01$) and found this does not change the overall results of this paper but does increase the uncertainty in our measurement of the \lirlp\ relation, as it limits the \iras\ sample to only 257 out of the 780 galaxies original galaxies (33\%). We therefore continue with the less restrictive $0.025 < z < 0.19$ selection to maximize the dynamic range of luminosities in our sample.
We find that this sample is $>$90\% complete above $>$340\,mJy and $>$81\% complete above $>$179\,mJy at 60\,\um.
\autoref{fig:z_distribs} shows the distribution of redshifts versus the IR luminosities for the \iras\ sample along with the other samples that we will describe below in $\S$\ref{subsec:hatlas_sample} and $\S$\ref{subsec:cosmos}.

Observational constraints included in the \iras\ sample are \wise\ 12\,\um, \iras\ 12\,\um, \wise\ 22\,\um, \iras\ 25\,\um, \iras\ 60\,\um, \iras\ 100\,\um, \hersch\ 100\,\um, \hersch\ 160\,\um, \hersch\ 250\,\um, \hersch\ 350\,\um, and \hersch\ 500\,\um\ with inclusion in the sample predicated only on a detection at 60\,\um.
These wavelengths allow us to probe approximately the full range of dust emission.
Because the 12\,\um\ band may capture PAH emission at the redshifts covered by this sample we test that the inclusion of two 12\,\um\ bands -- from both \iras\ and \wise\ -- does not bias our measurement of IR luminosities and peak wavelengths. We find that the median IR luminosities and peak wavelengths of galaxies in this sample differ by less than a percent when excluding either set of observations at 12\,\um. This difference is smaller than the average uncertainty in these parameters, so we conclude that the inclusion of both 12\,\um\ bands does not introduce additional bias.

To summarize, our selection function for this sample is galaxies that are (1) detected at ${\rm SNR} >5\sigma $ at 60\,\um\ in the \iras\ FSC, (2) are present in the H-ATLAS catalogs (${\rm SNR} >4\sigma $ at 250\,\um), (3) have a spec-$z$ in either RIFSCz or H-ATLAS, and (4) are in the redshift range $0.025 < z < 0.19$, taken to represent a snapshot in time of the local universe.
See \autoref{tab:selections} for a list of the wavelengths included and selection limits for this sample as well as the other samples described below.
\autoref{fig:det_lims_median_z} illustrates the detection limits for each sample in \lirlp\ space at their median redshifts with contours showing the distribution of sources. The goal of these selections is to create samples that are minimally biased with respect to SED dust temperature at a given IR luminosity. We will discuss this further in $\S$\ref{sec:discussion}.

Finally, in order to test that the \iras\ sample selection and observed filter set is not biased with respect to dust temperature we generate mock SEDs covering a grid in \lirlp\ space at peak wavelengths and IR luminosities that are broader than those we find in the sample, namely peak wavelengths between 35\,\um\ and 200\,\um, which corresponds to dust temperatures between 103\,K and 12.5\,K, respectively. We find no offset between the inputs to these noise-perturbed mock sources and the measured values for \lir\ and \lp\ at the signals to noise representative of the \iras\ sample with the wavelength sampling of the \iras\ sample.
In other words, over a range of mock SEDs that is more diverse than the actual observed \iras\ sample we find no bias due to the wavelength coverage or signal to noise of the \iras\ sample. If there were galaxies that populated parameter space in this grid we would be able to accurately measure their luminosities and peak wavelengths.
We return to a discussion of this test, with respect to any bias in this combination of filters, in $\S$\ref{sec:model_galaxies} after introducing the mock galaxy samples in more detail.

\subsection{H-ATLAS Sample}\label{subsec:hatlas_sample}
\begin{figure*}
    \centering
    \includegraphics[width=0.325\textwidth]{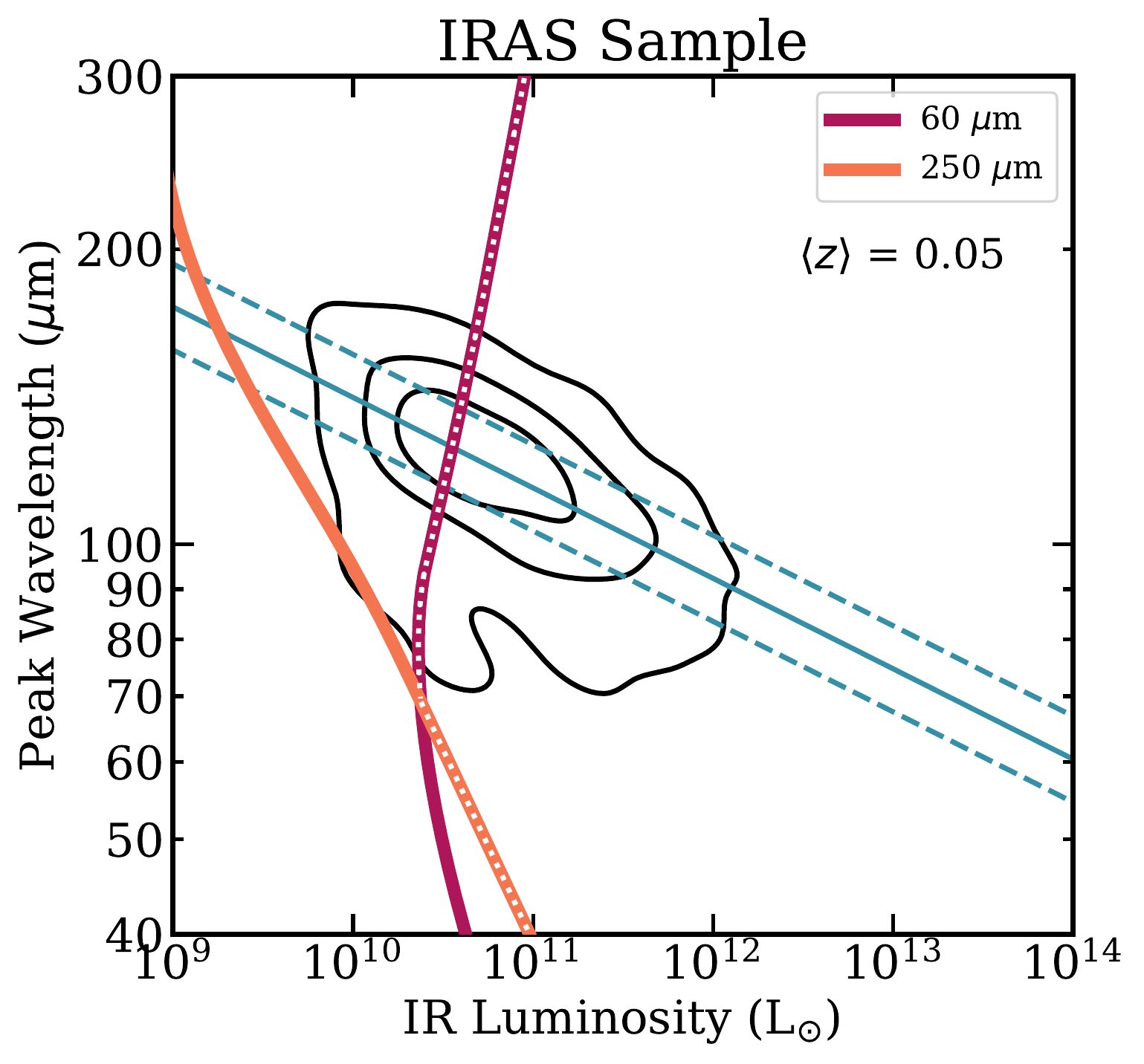}
    \includegraphics[width=0.325\textwidth]{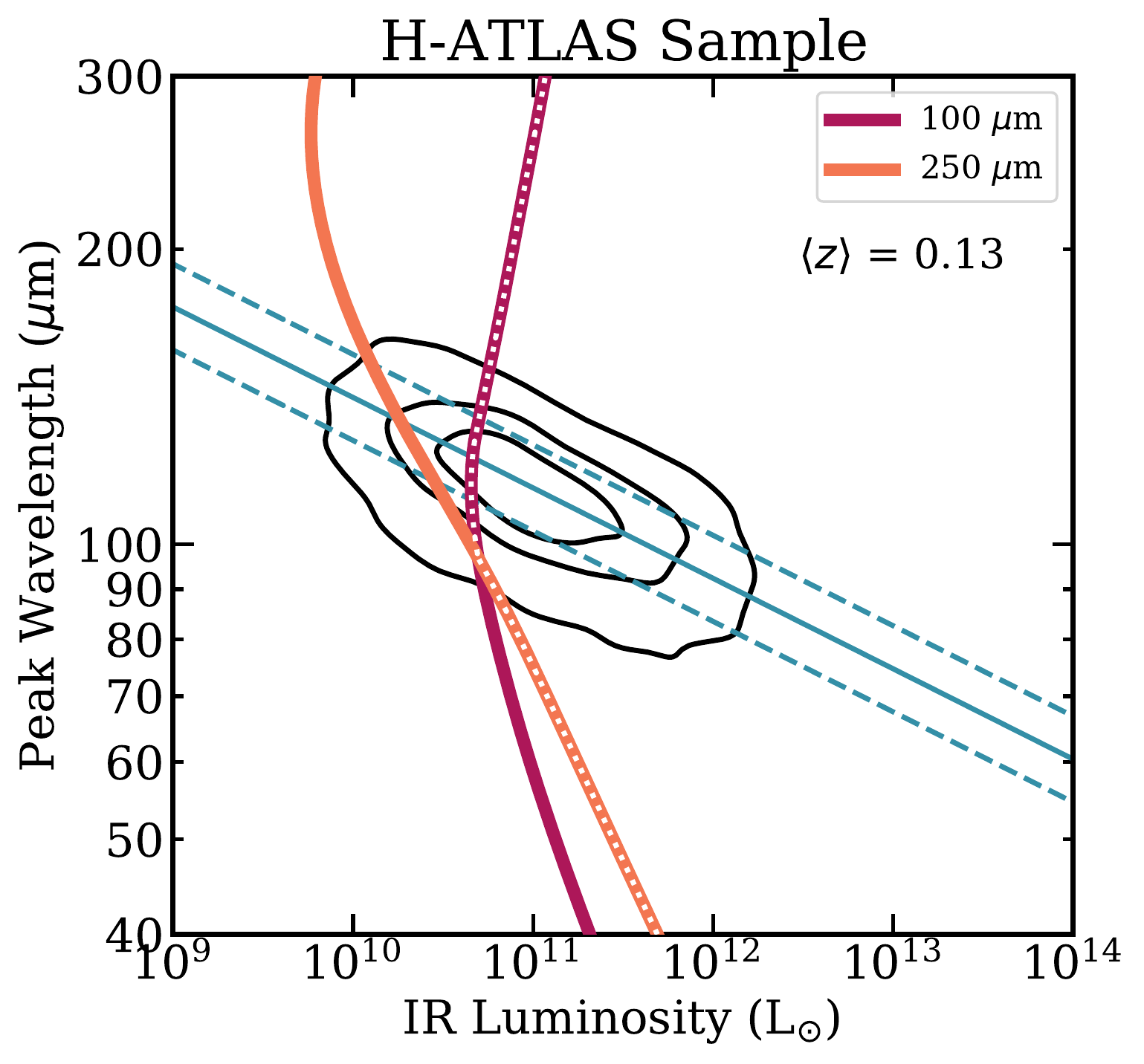}
    \includegraphics[width=0.325\textwidth]{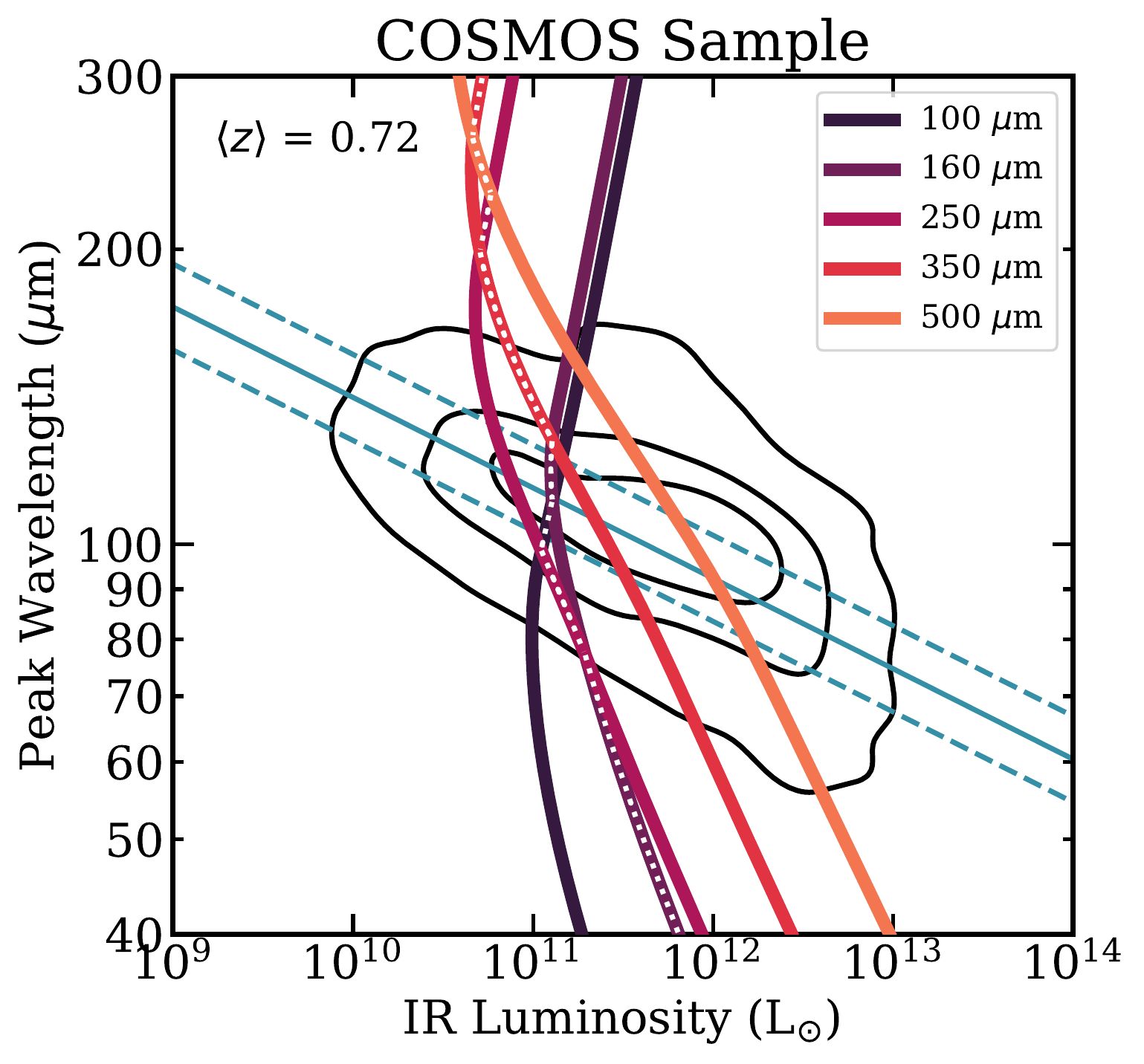}

    \caption{The detection limits at the median redshift for each sample overplotted on contours showing the density of detected galaxies for each sample. Contours from the outside to the inside depict the area containing 95, 68, and 34\% of the galaxies from each sample spanning their full redshift ranges.
    See \autoref{tab:selections} for the full quantified selection function of each of these samples. The solid light blue line shows the best fit to the \iras\ sample that is later described in $\S$\ref{sec:lirtd correlation}, and the dashed light blue lines show $\rpm$1$\sigma_{\lambda}$ of those fits. The white dotted lines show the net selection function that represents detection.}
    \label{fig:det_lims_median_z}
\end{figure*}

Our second sample, which we will refer to as the \hatl\ sample, expands the redshift horizon of this study beyond $z\sim0.2$, probing redshift evolution out to $z \sim 0.4$.
The sample is defined by galaxies detected at 250\,\um\ with SNR $> 4$ over an area of 161.6\,deg$^2$ from data release 1 of \hatl\ \citep[][]{Valiante16a, Bourne16a}.
For this sample, we make exclusive use of \hatl\ DR1 sources because \citeauthor{Bourne16a} perform a careful cross matching with \wise\ photometry \citep{Wright10a}, giving us important constraints on the sample's short wavelength dust SED.
There are of order 70,000 galaxies in \hatl\ DR1.
We further restrict our analysis to galaxies that are detected at 100\,\um\ with ${\rm SNR} > 3$, which significantly helps constrain the galaxies' \lp\ and reduces the sample to 3407 galaxies (5\% of all \hatl\ DR1 galaxies). The sky coverage with PACS and SPIRE in the \hatl\ DR1 is roughly the same, but the PACS instrument is much less sensitive than SPIRE. This accounts for the relative dearth of galaxies with SNR $>$ 3 at 100\,\um.
We will discuss the impact of this 100\,\um\ detection cut further in $\S$\ref{sec:model_galaxies}, but to summarize here, it effectively reduces the dynamic range in IR luminosity to the same range used for the \iras\ sample while also dramatically improving the SED fit quality by adding a precise constraint near the peak. In other words, the requirement of a 100\,\um\ detection removes low-\lir\ sources below the IR luminosity range studied in this manuscript.

Although H-ATLAS DR1 includes \wise\ observations at 12 and 22\,\um, we use only the 22\,\um\ band in our SED fits because PAH emission is expected to contribute to the 12\,\um\ band, depending on the redshift, and our modeling does not account for PAH emission.
In the \iras\ sample we included \wise\ 12\,\um, \iras\ 12\,\um, \wise\ 22\,\um, and \iras\ 25\,\um\ because we have as many as 5--6 constraints at wavelengths shorter than the peak wavelength of emission, so possible PAH contamination was downweighted by the other observations. On the short wavelength end the \hatl\ sample only has observations at 12\,\um, 22\,\um, and 100\,\um\ so PAH contamination in the 12\,\um\ band is more likely to skew our analysis.
While we do not require a detection with \wise, we do limit our sample to galaxies observed with \wise. Excluding those without \wise\ coverage removes 723 galaxies, reducing our sample to 2679 galaxies (by 22\%).
This sample covers $0 < z \lesssim 0.8$ but we limit our analysis to galaxies between $0.05 < z < 0.4$, due to a significant drop in \hatl\ survey sensitivity beyond $z\sim0.4$; i.e. \hatl\ is only sensitive to ULIRGs ($>$10$^{12}$\,\lsun) beyond $z \sim 0.4$, but their sky density at those redshifts is significantly lower, making average dust SEDs difficult to constrain.
As we did for the \iras\ sample, we use a lower redshift bound to avoid including spatially resolved sources. For this sample we choose a lower limit of $z = 0.05$ because this is the redshift where a galaxy with radius 10\,kpc is unresolved by \hersch\ at 250\,\um\ (i.e. within one beam). This lower limit reduces the number of galaxies to 2301, and the upper limit reduces the number of galaxies to 2167 (3\% of all \hatl\ DR1 galaxies).

\citet{Valiante16a} estimate the completeness of DR1 to be greater than 95\% for galaxies with flux densities above 42.15\,mJy at 250\,\um. This corresponds to 94\% of galaxies in our sample.
To address the possible inclusion of false positives, or galaxies whose far-IR emission is incorrectly matched to an optical/near-infrared source within the redshift range of interest at $z<0.4$, we evaluate the likelihood of random, non-physical associations in existing optical/near-infrared imaging in the H-ATLAS survey. We find the likelihood of a random match to an OIR source at the depth of SDSS DR10 data (22nd magnitude) at $z<0.4$ to be 0.7\% within a 3\arcsec\ matching radius, which would correspond to 15 sources.
We also investigate whether it is likely that there are sources present in \hatl\ whose optical counterparts are not detectable in the available optical imaging from SDSS DR10\footnote{Both the \iras\ and COSMOS samples have deeper relative OIR imaging (\iras\ is also matched to SDSS DR10, but sources are at lower redshift than the \hatl\ sample, and the COSMOS OIR imaging is much deeper than that of SDSS DR10 with average broad band depths $\sim$26--27 AB), so we can conclude that all galaxies in each sample have detectable OIR counterparts in their corresponding optical imaging surveys.}. The infrared excess, the ratio of the IR to the UV luminosities (${\rm L_{\rm IR}}/{\rm L_{\rm UV}} \equiv {\rm IRX}$) of sources at $z=0.4$ that have U band flux densities at the detectability limit of SDSS DR10 (AB magnitude 22) and \hersch\ 250\,\um\ fluxes at the detectability limit of \hatl\ correspond to ${\rm IRX} \sim 4\times10^{5}$, which is much higher than a more typical value of IRX for DSFGs in the literature $\sim 10^3$--$10^{4}$ \citep[e.g.][]{Howell10s, Casey14a}.
Adopting an IRX value of $10^4$ implies an AB magnitude limit of 18, implying all OIR counterparts should be detected in SDSS DR10.

The included wavelengths of observation for the \hatl\ sample are then 22\,\um, 100\,\um, 160\,\um, 250\,\um, 350\,\um, and 500\,\um.
See \autoref{fig:z_distribs} for the redshift distribution versus IR luminosity of the H-ATLAS sample. \autoref{fig:det_lims_median_z} shows the detection limits at the median redshift of the \hatl\ sample with contours showing the distribution of sources in \lirlp\ space.
To summarize, our selection function for this sample is galaxies that have (1) SNR $>$ 4 at 250\,\um\ and SNR $>$ 3 at 100\,\um\ in \hatl\ DR1, (2) are observed with \wise\ (though no detection is necessary), and (3) lie between $0.05 < z < 0.4$.

\subsection{COSMOS Sample}\label{subsec:cosmos}
Our third sample, which we will refer to as the COSMOS sample, extends out to $z=2$. It contains \hersch-selected galaxies from the COSMOS field \citep{Scoville07a} originally presented in \citeauthor{Lee13a} (\citeyear{Lee13a}; hereafter L13).
L13 measures \spitz\ MIPS flux densities from \citet{Sanders07c}, \hersch\ SPIRE and PACS flux densities from \citet{Oliver12a, Lutz11a}, 1.1\,mm AzTEC \citep{Scott08c, Aretxaga11p}, and 1.2\,mm MAMBO \citep{Bertoldi07f} flux densities. They use positional priors from 24\,\um\ \citep{lefloch09a} and 1.4\,GHz \citep{Schinnerer07w} following the XID linear inversion technique \citep{Roseboom10a, Roseboom12a}.
The authors then match the 24\,\um\ positions to $K_{s}$-band counterparts from the COSMOS-WIRCam near-infrared imaging survey \citep{McCracken10a} using a 2\arcsec\ search radius.
Finally, these $K_s$-band positions are matched with a 1\arcsec\ search radius to the photometric redshift catalog of \citet{ilbert09a} to derive photometric redshift constraints.
This results in a total of 39,333 sources. Most of these are not significantly detected in the IR/mm and so are excluded from the analysis in this paper upon applying further selection cuts.

In the present paper we update the L13 sample and methodology using more recent COSMOS data releases.
We start with the 4220 galaxies in the L13 catalog with SNR $>$ 3 in at least two PACS or SPIRE bands (11\% of all L13 galaxies). 
L13 uses only galaxies with SNR $>$ 5 in at least two PACS or SPIRE bands (1810 galaxies, using confusion limits as a floor on the errors on flux density). We explored both SNR cuts for our final sample and found that pushing down to SNR $>$ 3 added significant dynamical range to the luminosity limit of our analysis and does not detract from the quality of our SED constraints.
We then cross match L13 positions for these 4220 galaxies with ALMA sources in the A$^3$COSMOS catalog \citep{Liu19e} given the significant uncertainty in the initial 24\,\um/1.4\,GHz association with \hersch\ flux densities.
\citet{Magdis11q} has shown that 98\% of \hersch\ sources at $z < 2$ will be detected in deep \spitz\ 24\,\um\ imaging. Similarly, ALMA detections centered on \hersch-selected sources are physically associated due to the rarity of both populations on the sky. Therefore the likelihood that an ALMA source located within the MIPS 24\,\um\ point spread function radius is not associated with the galaxy responsible for 24\,\um\ emission at $z < 2$ is low ($<$1\%).
A$^3$COSMOS contains all available ALMA archival continuum pointings in the COSMOS field with detections of 756 unique galaxies across $\sim$280\,arcmin$^2$.
We use a search radius of 3$\arcsec$, the beam radius of the \spitz\ 24\,\um\ band, to match 24\,\um\ positions to ALMA-detected counterparts.
ALMA detections were found for 147 sources in L13 (4\% of the original 4220 sources).
Of these, we find 20 sources (14\% of the 147 matches) were misidentified in the L13 sample based on the adopted positional prior. In other words, the original OIR counterpart identified within the 24\,\um\ beam by L13 was found to be spatially offset from an ALMA source, also within the 24\,\um\ beam, by larger than 20\,kpc for 14\% of the subsample with ALMA measurements available. These mismatches did not affect the photometry of these sources, as the XID linear inversion technique simply measures the fluxes in all bands at the positions of 24\,\um\ \spitz\ sources. A new matched position within the beam of 24\,\um\ emission only results in a new matched OIR counterpart and hence a new redshift.

After adopting the new positions from the A$^3$COSMOS subset, we cross match all source positions (including those without ALMA data) with the COSMOS2020 photometric redshift catalogs \citep{Weaver21j}.
Their photometric redshifts are updated to make use of the near-IR and optical data from the UltraVISTA and Subaru HSC surveys' deeper imaging in the field.
The COSMOS2020 phot-$z$ catalog consists of two separate photometric catalogs generated using different source extraction techniques as well as two photometric redsshift (phot-$z$) fitting techniques applied to those two catalogs, resulting in four unique phot-$z$ catalogs. See \citet{Weaver21j} for more details.
Following the procedure adopted by L13 and \citet{lefloch09a}, we perform our spatial cross match from long wavelength astrometry (ALMA if available, otherwise \spitz\ 24\,\um) separately on all COSMOS2020 catalogs using a 2\arcsec\ search radius, and adopt the nearest neighbor match. These works found that by adopting a 2\arcsec\ search radius they were able to reduce the fraction of multiple matches while still taking into account most of the positional uncertainty associated with the 3\arcsec\ width of the MIPS 24\,\um\ PSF. For sources with ALMA positions, we adopt a search radius of 1\arcsec, as this is the size of the average ALMA beam in A$^3$COSMOS.
Thus, from COSMOS2020 we have between 1--4 photometric redshift estimates for each galaxy.
There are 4011 galaxies matched out of the original 4220 sources with SNR $>$ 3 in at least two \hersch\ bands (95\%).
Next we remove galaxies that are marked as masked in both source catalogs (for proximity to a bright star which obfuscates the OIR source photometry), leaving only 2295 galaxies of the original 4011 with matches (57\%).
Rather than fitting these masked galaxies with the original L13 phot-$z$s we choose to only retain the subset of sources matched to COSMOS2020 to retain consistency in how redshifts are measured within the sample. We do not find that this cut hinders our final analysis in any way, as the remaining sample is still sufficiently large.

Where spectroscopic redshifts are available in the COSMOS2020 catalog we use them instead of photometric redshifts. Of the 2295 sources, 285 have spec-$z$s (12\%).
For the remaining galaxies we adopt a single phot-$z$.
For those with two phot-$z$ estimates from the COSMOS2020 catalogs we use the lower $z$ because galaxies in this sample are statistically more likely to be at lower $z$.
For galaxies with three phot-$z$ estimates we adopt the median redshift.
And finally, for galaxies with four phot-$z$ estimates we adopt the redshift that is the closest to the median. We do this rather than adopting the true median so that we may retain the redshift uncertainties included with the COSMOS2020 catalogs.
There are 2005 sources with 4 phot-$z$ estimates, 83 with 3 phot-$z$ estimates, 206 with 2 phot-$z$ estimates, and 1 source with 1 phot-$z$ estimate. The median standard deviation of phot-$z$ estimates is 0.02 for sources with 4 estimates.
Note that only a minority of sources (423 of 2295, or 18\%) have standard deviations in their 2--4 photometric redshift estimates greater than 0.1. Additionally, the majority of our galaxies have very similar photometric redshift estimates to those from L13, with $\sigma_{\Delta z / (1+z)} = 0.009$.
See \autoref{fig:z_distribs} for the distribution of COSMOS sample redshifts and IR luminosities relative to the other samples.

\begin{figure*}
    \centering
    \includegraphics[width=0.32\linewidth]{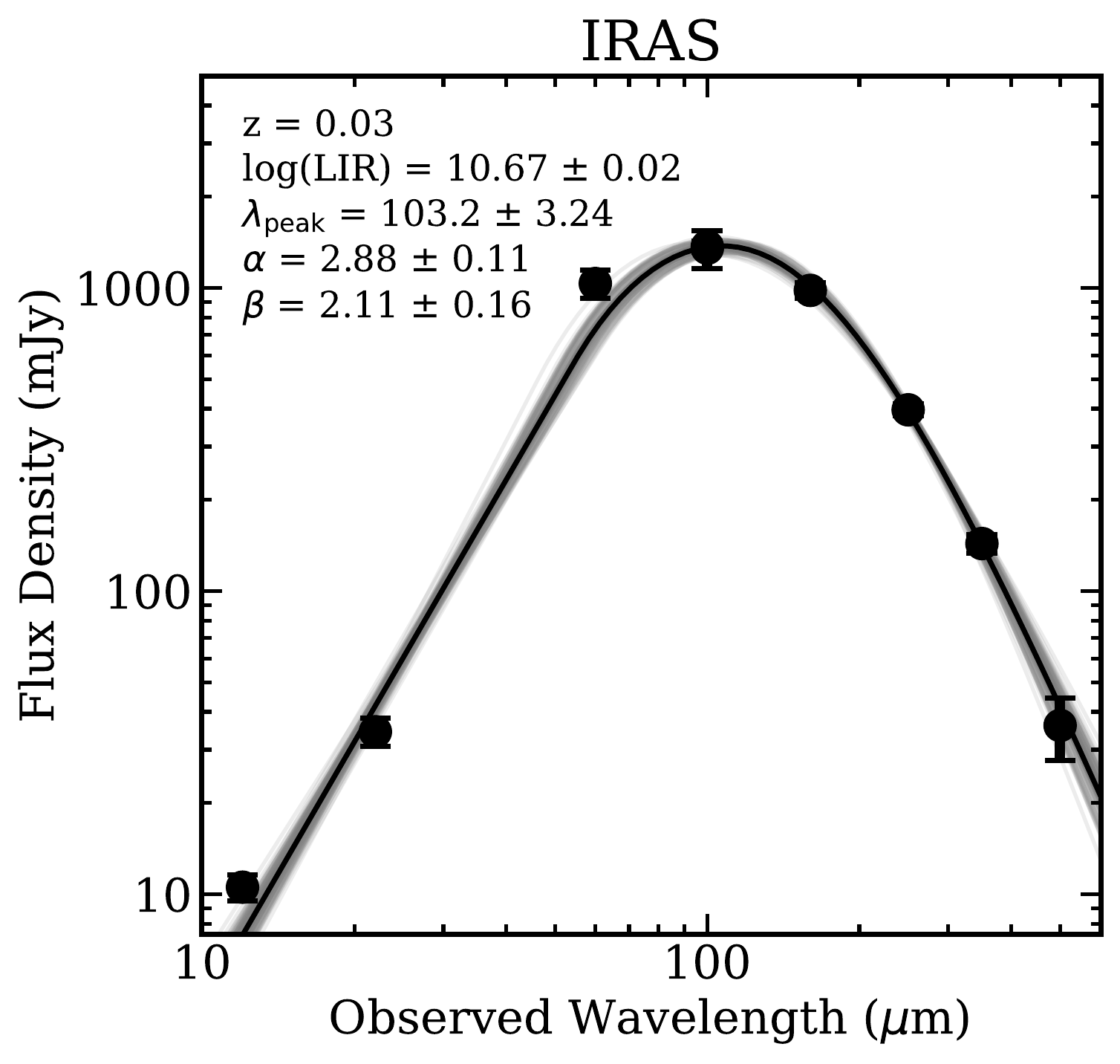}
    \includegraphics[width=0.32\linewidth]{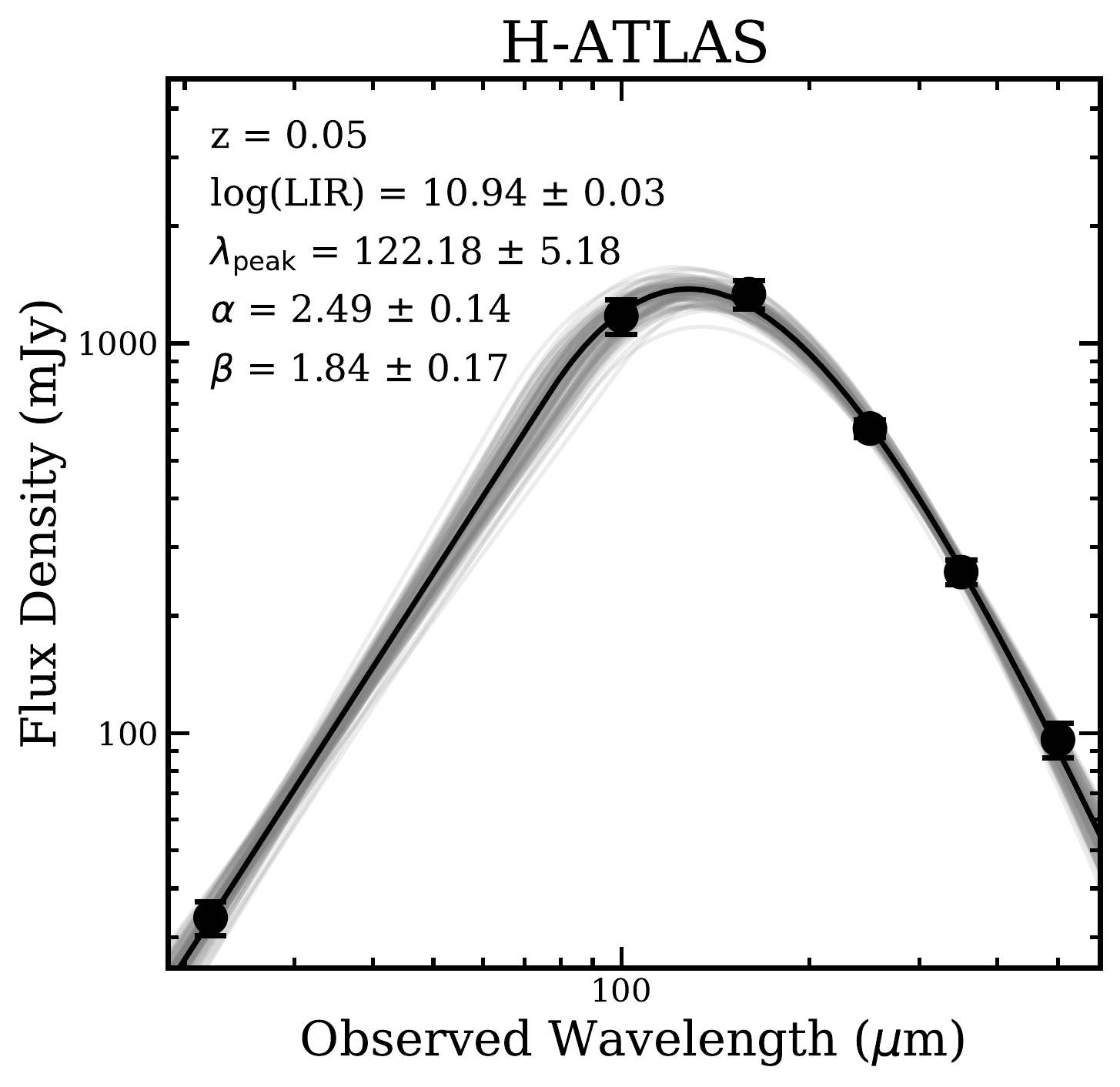}
    \includegraphics[width=0.32\linewidth]{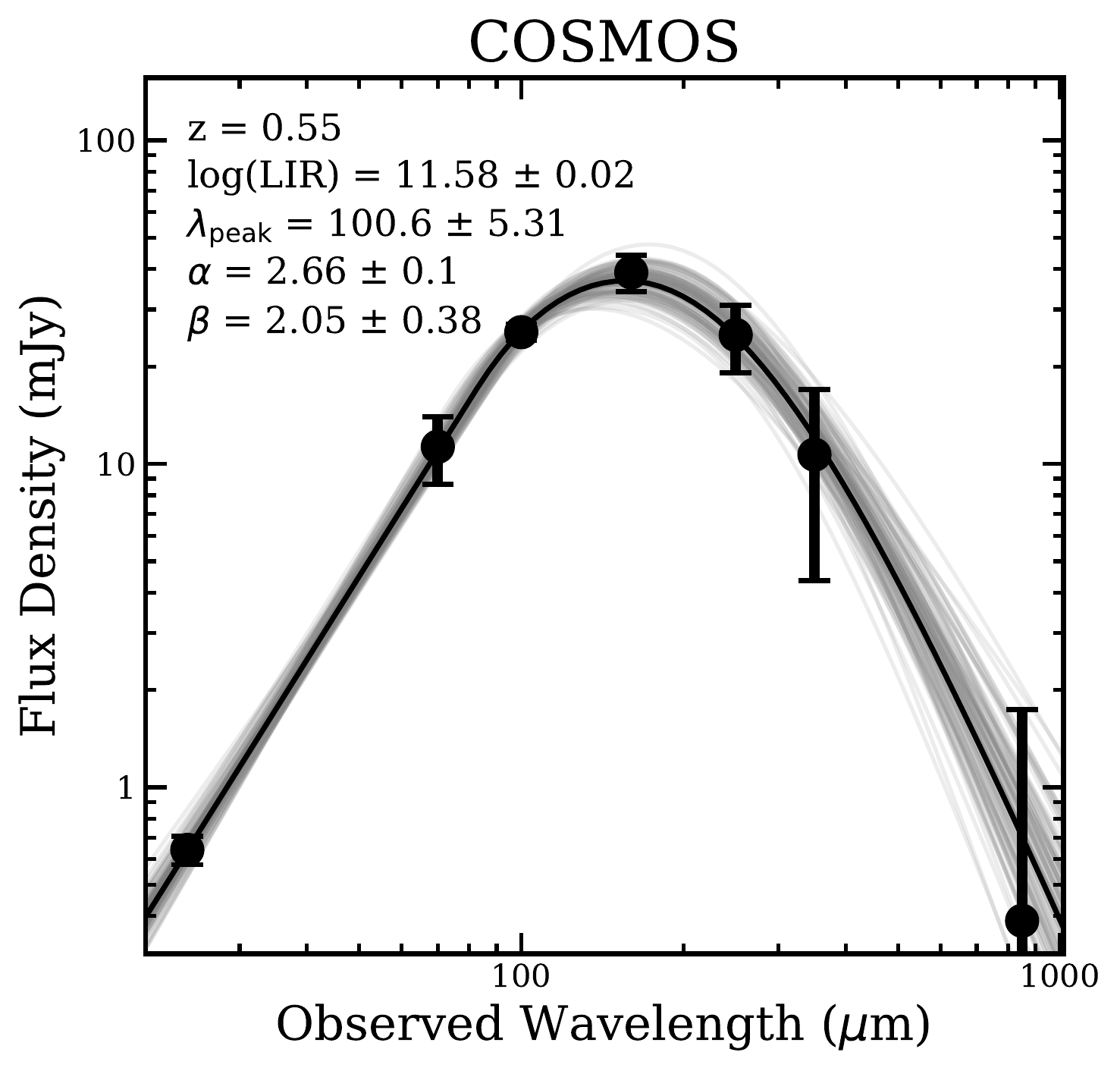}
    \includegraphics[width=0.32\linewidth]{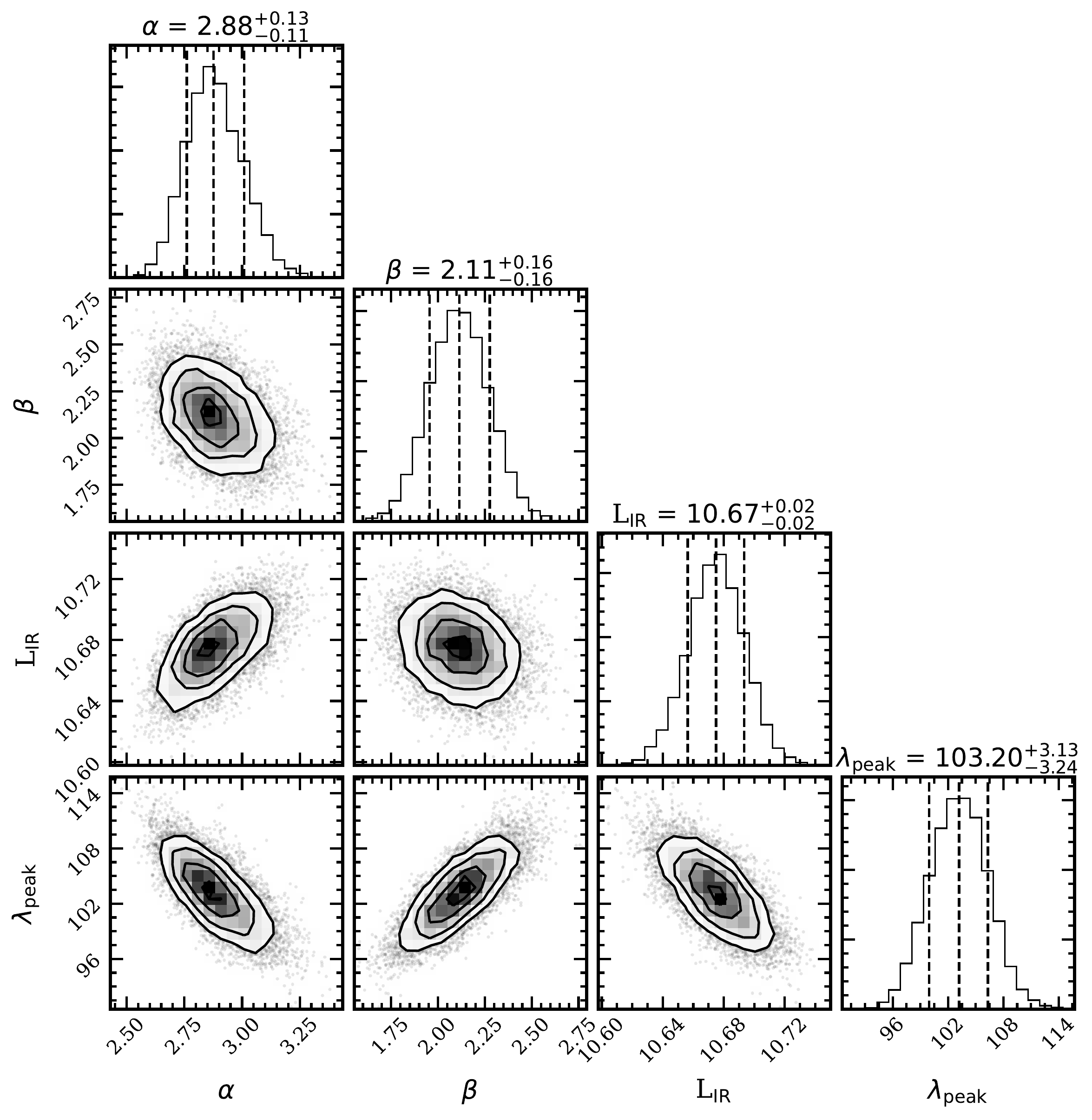}
    \includegraphics[width=0.32\linewidth]{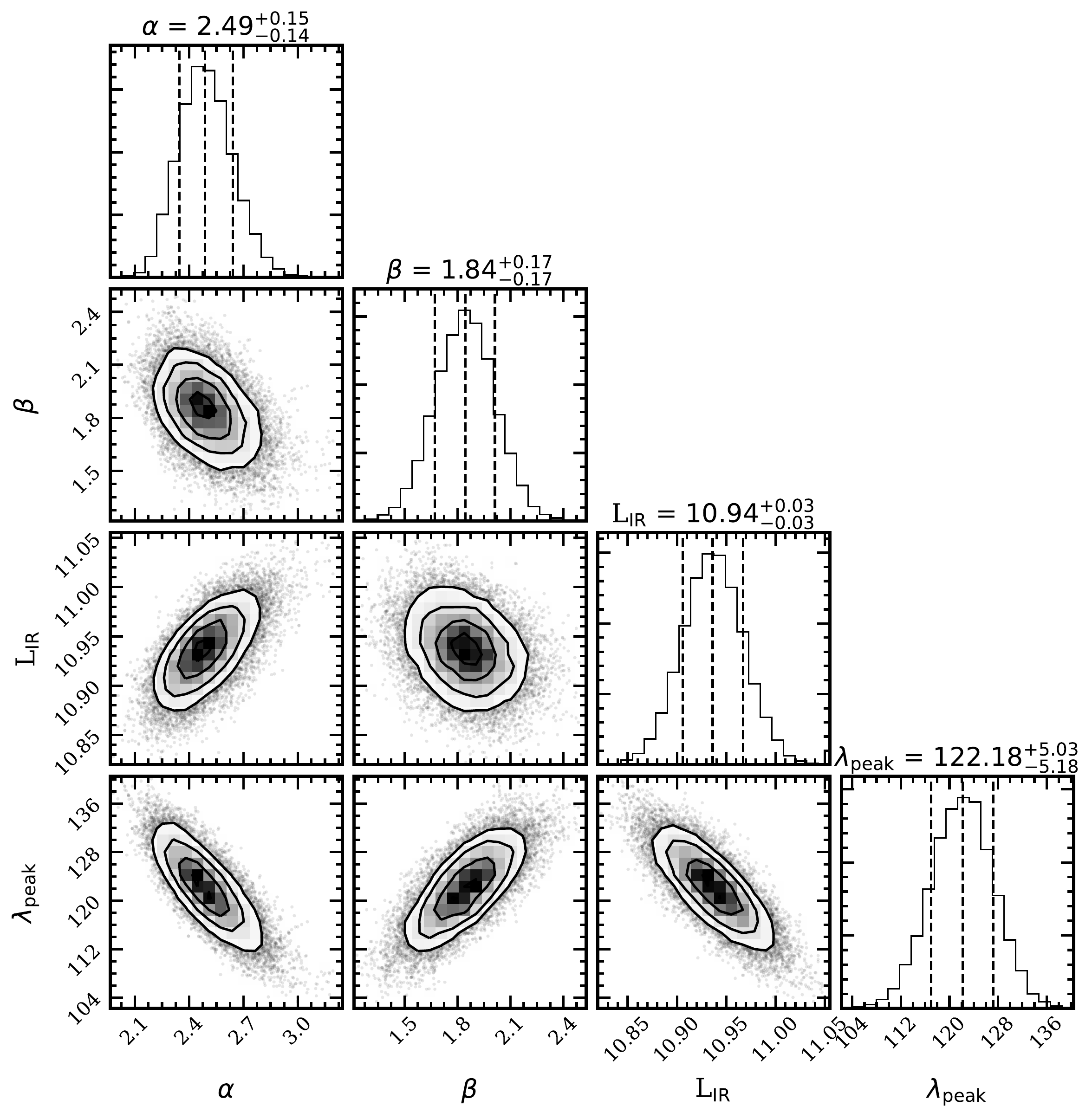}
    \includegraphics[width=0.32\linewidth]{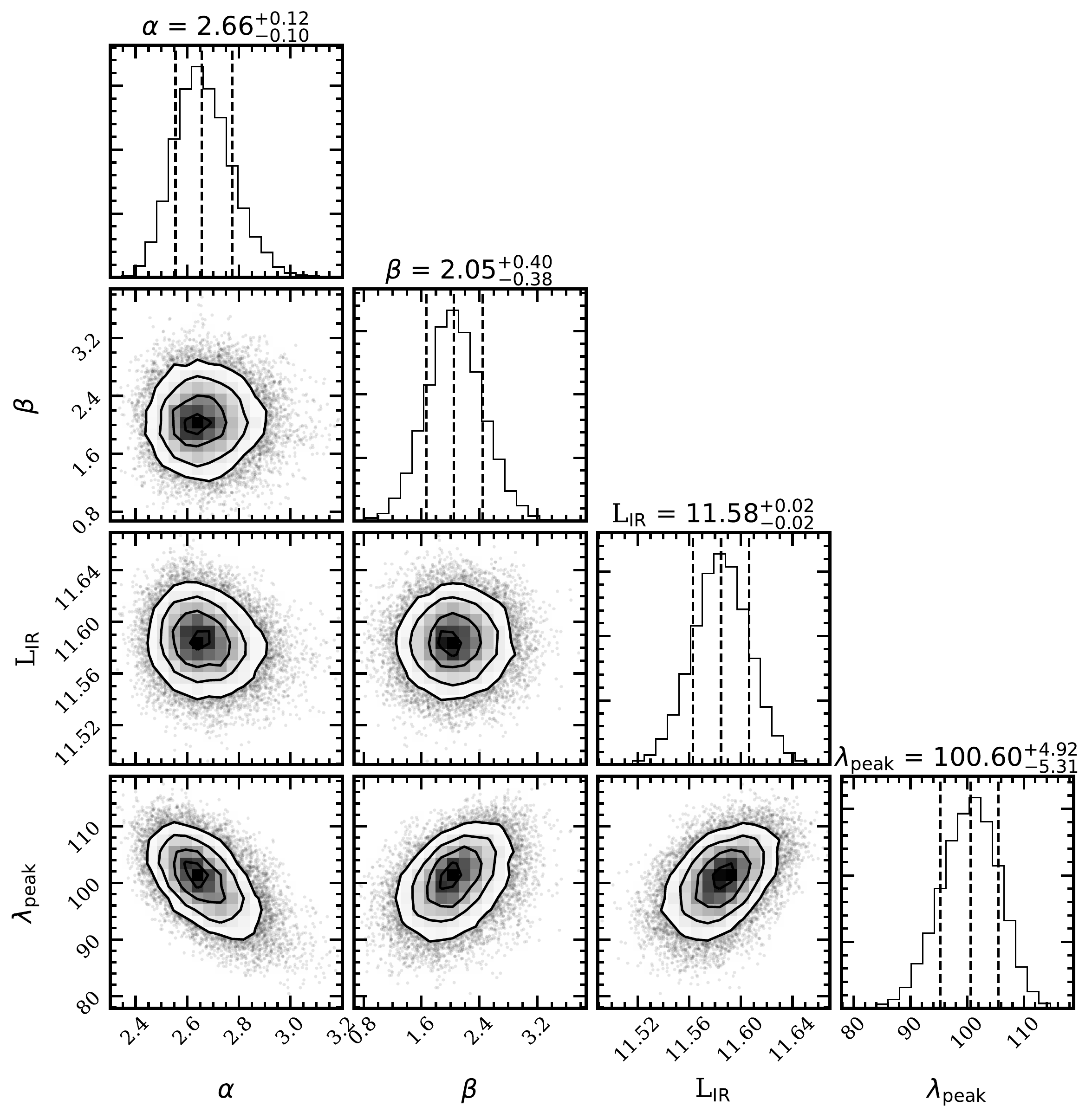}
    \caption{Example SED fits to galaxies from each of our three samples. The black line on the SED is the best fit curve and the grey lines are 50 randomly selected MCMC chains. The SED fits are shown at top and the two dimensional joint posterior distributions for fit parameters are shown at bottom. The posterior distributions show mild correlation between parameters. We fix $\lambda_0=200$\,\um\ in this work. For galaxies across all samples, typical errors on $\alpha$ are 7\%, on $\beta$ are 20\%, on \lir\ are 10\%, and on \lp\ are 9\%.}
    \label{fig:fit_and_corner}
\end{figure*}

Galaxies in the COSMOS sample span $0.02 < z < 4.8$. We limit our analysis to $0.15 < z < 2$ due to the poor sensitivity of \hersch\ for IR-luminous sources beyond $z\sim2$ and potential bias in the dust temperature measured from those SEDs beyond that epoch. 
Unlike the \hatl\ sample where a lower redshift bound of 0.05 is chosen to exclude galaxies resolved with \hersch, for the COSMOS sample we increase our lower redshift bound to 0.15 because the COSMOS field is much smaller than that of \hatl, and a small area provides a low number of sources per given bin of $\Delta z$ below $z < 0.15$. Whereas there are 72 sources between $0.05 < z < 0.15$ in the COSMOS sample, there are 1314 sources in the \hatl\ sample over this same range. 
Low number statistics between $0.05 < z < 0.15$ are likely to result in a bias to our analysis, particularly given the lower IR luminosities of that subsample, so we require $>$100 galaxies per $\Delta z = 0.1$ bin on the low redshift end of the COSMOS sample. We achieve this by adopting a lower redshift bound of 0.15. This reduction takes the COSMOS sample to 1990 sources.

The last step in our sample selection procedure is to cross match the COSMOS sample with sources from the S2COSMOS catalog \citep{Simpson19a} using a search radius of 6\farcs5, the beam radius of SCUBA-2 at 850\,\um, in order to fold in their 850\,\um\ data to our SED analysis. This results in 124 detections with SNR $>$ 3, and 9 with SNR $>$ 5. 
For galaxies in the COSMOS sample that do not have a matched S2COSMOS source we adopt the flux in the S2COSMOS 850\,\um\ map at the location of our source. For the uncertainty on this flux density we adopt the RMS from the S2COSMOS 850\,\um\ RMS map. 
The wavelengths of observation for our COSMOS sample are then 24, 70, 100, 160, 250, 350, 500, 850, 1100, and 1200\,\um.
See \autoref{fig:z_distribs} for the redshift distribution of sources in the COSMOS sample versus IR luminosity and see \autoref{fig:det_lims_median_z} for the detection limits at the median redshift of the COSMOS sample with contours showing the distribution of sources in lirlp space.
We find that this sample is $>$90\% complete at flux densities $>$10\,mJy at 250\,\um.
To summarize, our selection function for this sample is galaxies from L13 that have (1) SNR $>$ 3 in at least two of five \hersch\ bands (100, 160, 250, 350, 500\,\um), and (2) have redshifts between $0.15 < z < 2.0$.

\section{Modeling galaxies' IR SEDs}
\label{sec:sed_model}

To model the SEDs of galaxies between 8 and 1000\,$\mu$m we use a piecewise function consisting of a mid-infrared power law and a far infrared modified blackbody. The model takes the form:
\begin{equation}\label{eq:1}
  S(\lambda) =
      \begin{cases}
        N_{pl}\lambda^{\alpha} & \text{: } \frac{\partial\log S}{\partial\log \lambda} > \alpha \\
        
        \frac{N_{bb} \left(1-e^{-(\lambda_{0}/\lambda)^{\beta}}\right) \lambda^{-3}}{e^{hc/\lambda kT}-1} & \text{: } \frac{\partial\log S}{\partial\log \lambda} \leq \alpha,
      \end{cases}
\end{equation}
where $\lambda$ is the rest-frame wavelength, $\alpha$ is the mid-infrared power law slope, $N_{pl}$ and $N_{bb}$ are normalization constants, $\lambda_0$ is the wavelength where the dust opacity equals unity, $\beta$ is the dust emissivity index, $T$ is the luminosity-weighted characteristic dust temperature, $h$ is the planck constant, $k$ is Boltzmann constant, and $c$ is the speed of light. The two piecewise functions connect seamlessly at the wavelength where the slope of the modified blackbody is equal to the slope of the power law function.

To fit this function to the data presented herein we built a package called Monte Carlo InfraRed SED or MCIRSED. It is described in further detail in the Appendix, $\S$\ref{sec:bayesian_modeling}. To summarize, we use Markov chain Monte Carlo (MCMC) to sample the posteriors of our fit parameters and provide confidence intervals.
The fit parameters that may be fixed or free are the power law slope, $\alpha$, the dust emissivity index, $\beta$, and the wavelength where the opacity equals unity, $\lambda_0$. The peak wavelength and IR luminosity are always free parameters. 
\autoref{fig:fit_and_corner} shows example fits to a galaxy from each of our samples with their associated two dimensional joint probability distributions. The only parameter that is fixed for all galaxies in our analysis is $\lambda_0 = 200$\,\um, consistent with measurements or assumptions made for similarly IR-luminous galaxies in the literature \citep[e.g.][]{Blain03a, Draine06a, Conley11a, Rangwala11d, Greve12a, Conroy13a, Spilker16a, Simpson17a, Zavala18a, Casey19a}.

For our fiducial \iras\ sample we use flat bounded priors on $\alpha$, $\beta$, and \td. The bounds are chosen to restrict sampling to parameter space that corresponds to conditions that are physical though not restrictive. For the \hatl\ and COSMOS samples we use normally distributed priors based on the best fits to the \iras\ sample. Details on the choice of parameter priors are given in \autoref{sec:bayesian_modeling}. 
When fitting galaxies with upper limits in any of the wavelength bands with unspecified measurement uncertainty we adopt a flux density of 0.0 with an uncertainty equal to the implied 1$\sigma$ upper limit of the given dataset. However, if a band has a flux density and uncertainty reported in the parent catalog we fit with those reported values, even if the SNR of the observation is low or negative. In practice, measurements with negative SNR (originating from negative flux density and positive uncertainty) are limited to SNRs close to zero, allowing fits to remain exclusively at positive flux densities and still within the typical uncertainty on the measurement.
The best fit parameters reported throughout this work are the median values of the posterior distributions and the uncertainties are the 16th and 84th percentiles ($\rpm$1$\sigma$ for Gaussian distributions).

\section{Measuring the \lirlp\ correlation}\label{sec:lirtd correlation}

\begin{figure*}
    \centering
    \includegraphics[width=0.49\textwidth]{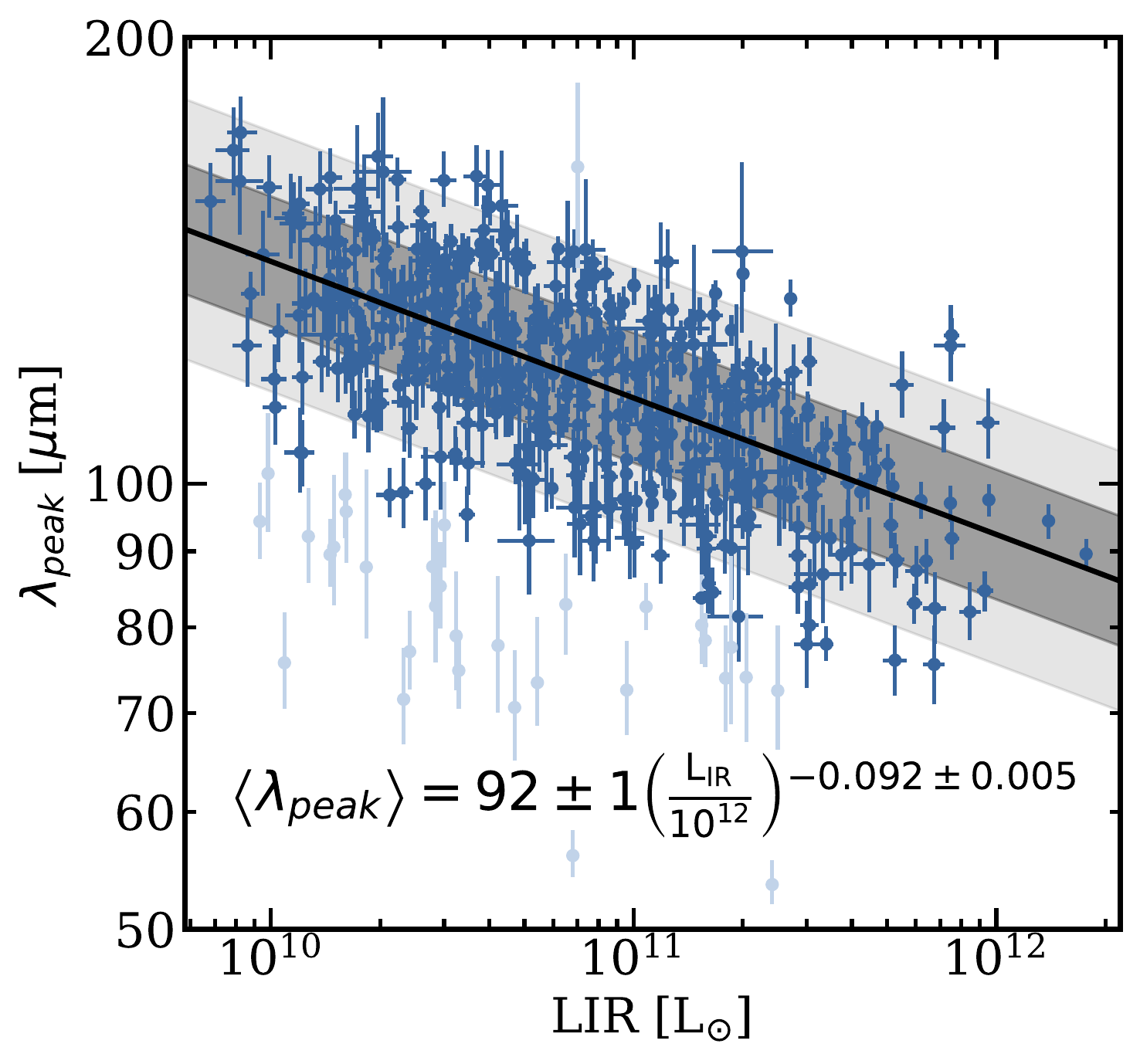}
    \includegraphics[width=0.417\textwidth]{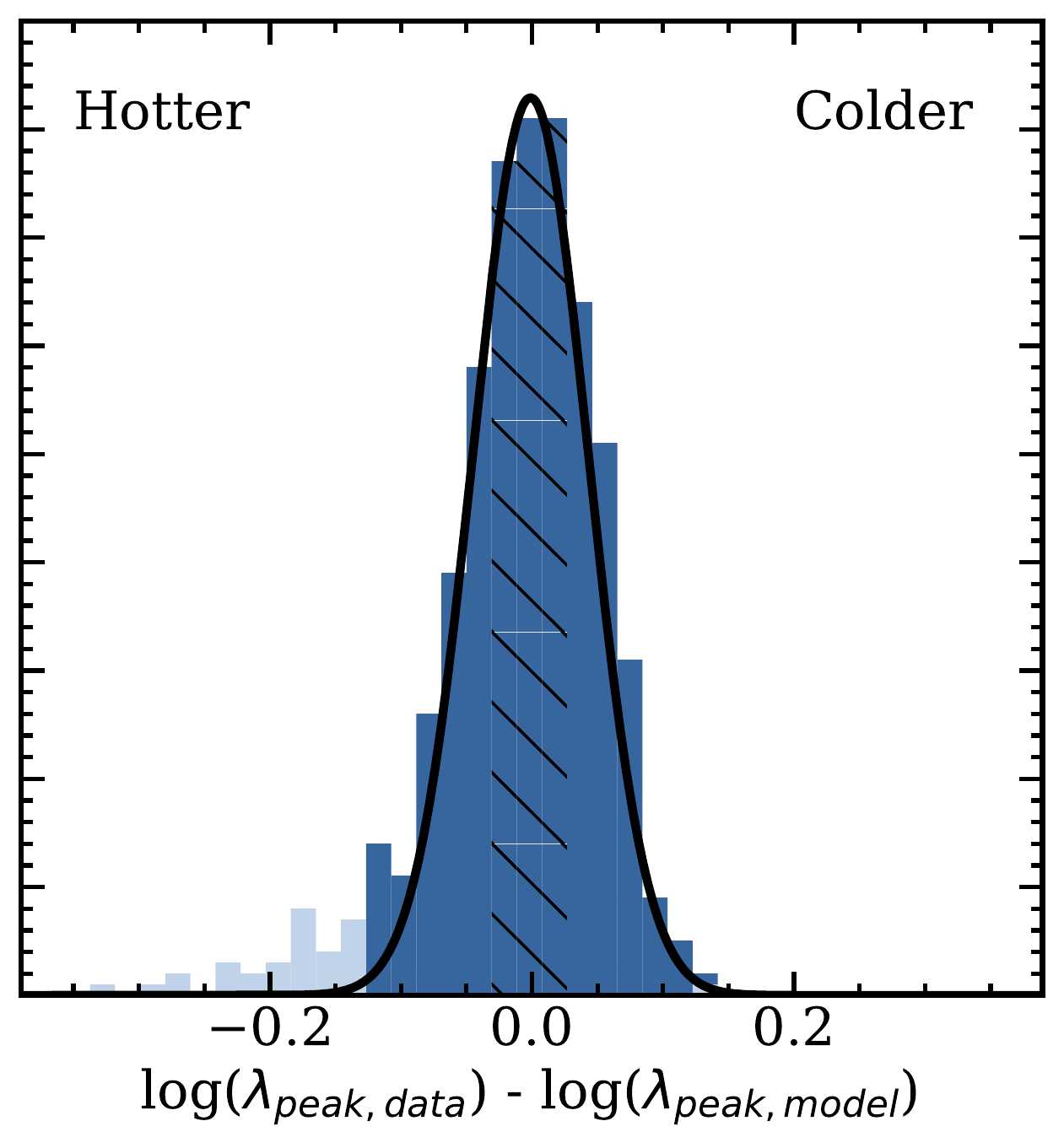}
    \caption{Left: The \lirlp\ correlation for the \iras\ sample. The black line is the best fit to the correlation using \autoref{eq:lirlp_corr}.
    The darker grey shaded region denotes $\rpm$1 standard deviation of the distribution ($\sigma_{\lambda}$) of points around the best fit line. The lighter shaded region denotes $\rpm$2 standard deviations. In order to fit the correlation in the presence of outliers we use iterative sigma clipping while fitting with a cut threshold of $>$3\,$\sigma$.
    Outliers are shown in lighter blue points.
    Independent matching to catalogs of luminous AGN indicate a higher percentage of outliers are AGN than those on the relation (20\% of outliers vs 6\% of inliers).
    Right: Histogram of log(data) - log(model) for the \lirlp\ correlation fit to the \iras\ sample. As in the left panel, sigma clipped outliers are marked in lighter blue.}
    \label{fig:IRAS_lirtd}
\end{figure*}

Our investigation into whether there is redshift evolution in the bulk SED characteristics of dust in galaxies between $0 < z < 2$ is anchored to our analysis of the empirically established correlation between \lir\ and \lp.
We choose \lir\ as the fundamental quantity to which we compare dust temperatures, via \lp, for two reasons: 1) IR luminosity is the most fundamental quantity we can derive directly from a galaxy's dust SED other than \lp, making it directly measurable for all galaxies in our samples, 2) it has been shown to be more closely correlated with galaxy-integrated dust temperature \citep{Burnham21i} than several alternate quantites like stellar mass, specific star formation rate, or distance from the ``main sequence'' (c.f. \citeauthor{Chapman03b} \citeyear{Chapman03b}; \citeauthor{Chanial07e} \citeyear{Chanial07e}; \citeauthor{Symeonidis13a} \citeyear{Symeonidis13a}; \citeauthor{Magnelli14a} \citeyear{Magnelli14a}; \citeauthor{Lutz16c} \citeyear{Lutz16c}). 
Though the star-formation surface density, $\Sigma_{\rm IR}$, seems to be more fundamentally correlated with dust temperature than \lir\ (as one might suspect given its dependence on dust geometry; e.g. \citeauthor{Lutz16c} \citeyear{Lutz16c}; \citeauthor{Burnham21i} \citeyear{Burnham21i}), the lack of high-resolution infrared imaging, thus measurement of dust sizes, for the vast majority of our galaxies prevents direct analysis of possible evolution in $\Sigma_{\rm IR}$ vs \lp. Thus, we anchor our work to the \lir\ vs. \lp\ plane, as the next best alternate for unresolved sources, for the rest of this paper.

The correlation between IR luminosity and dust temperature (where \td\ $\propto$ 1/\lp) has been well documented in the literature.
Early studies of \iras\ galaxies in the local universe demonstrate that IR color, a parameterization of dust temperature, correlates with IR luminosity \citep[e.g.][]{Andreani96i, Dunne00w, Dale01a, Chapman03b}.
The launch of \hersch\ revealed galaxies at higher redshifts also obeyed a correlation between dust temperature and luminosity \citep[e.g.][]{Hwang10a, Magnelli12a, Casey12b, Casey12c, Symeonidis13a, Casey18a}.

The specific hypothesis we are testing is that the correlation between IR luminosity and peak wavelength does not evolve with redshift. This is suggested to be the case in some recent papers \citep{Casey18a, Reuter20a}, though \citet{Casey18a} points out their analysis has limited power to constrain the relation at $z > 2$ because of how few galaxies are in their samples at these redshifts. Similarly, the sample presented by \citet{Reuter20a} is limited to $\sim$100 sources over a range of redshifts between $1.9 < z < 6.9$.
Establishing whether there is redshift evolution in the \lirlp\ correlation, especially using a uniform approach to SED fitting is important because it might imply a corresponding evolution in the average properties of dust in the ISM of galaxies across these redshifts.

To model the \lirlp\ correlation we begin with the equation from \citet{Casey18a} that relates the peak wavelength to the \lir\ of galaxies:

\begin{gather}
    \label{eq:lirtd}
    \left< \lambda_{peak}({\rm L_{\rm IR}})\right> = \lambda_{\rm t} \left(\frac{\rm L_{\rm IR}}{\rm L_{\rm t}}\right)^{\eta}.
\end{gather}
Here \lt\ is fixed to 10$^{12}$\,\lsun, $\lambda_{t}$ is the peak wavelength of galaxies at \lt, and $\eta$ is the power law index.
Note that \lt\ is simply a normalization constant and is fixed to 10$^{12}$\,\lsun\ as is done in \citet{Casey18a}; its adopted value does not impact the measured results.
Because we find the data in each of our samples is normally distributed in log(\lir)--log(\lp) space we also model the distribution of galaxies about this best-fit line as a Gaussian using the general prescription of \citet{Kelly07a}. Our model then becomes:

\begin{gather}
    \label{eq:lirlp_corr}
    \log(\lambda_{\rm peak}) = \log(\lambda_{\rm t}) + \eta [\log(\rm L_{\rm IR}) - 12] + N(\mu, \sigma_{\log\lambda_{\rm peak}})
\end{gather}
where $N(\mu, \sigma_{\log\lambda_{\rm peak}})$ is a Gaussian with mean $\mu \equiv 0$, and standard deviation $\sigma_{\log\lambda_{\rm peak}}$, which we will refer to as $\sigma_{\lambda}$ for convenience.
Also for simplicity, we will hereafter refer to the set of all fit parameters as $\Theta = (\lambda_t, \eta, \sigma_{\lambda})$. 

Each sample is fit using MCMC with non-informative priors on all parameters to derive independent measurements of $\Theta$.
The rest of this section presents measurements for several sets of galaxies, starting with the \iras\ sample, which we use as a basis, indicative of the relationship in the local universe at $z < 0.2$. We then proceed to simulate mock galaxies, adopting the derived relation from the \iras\ sample. The mock galaxies are used to infer the influence of telescope selection biases on IR-luminous galaxy samples, which become more prominent at higher redshifts. Lastly, we present fits to both the \hatl\ and COSMOS samples and compare those fits to the expectations from the mock galaxy samples passed through the same observational selection filters.

%%%%%%%%%%%%%%%%%%%%%%%%%%%%%%%%%%%%%%%%%%%%%%%%%%%%%%%%%%%%%%%%%%%%
\subsection{Fitting the $z < 0.2$ \iras\ Sample as an Anchor}
\begin{table*}[]
    \caption{The best fit parameters of the \lirlp\ correlation for each redshift bin across all samples. Each bin is consistent with the case of no redshift evolution since the redshift range of the \iras\ sample, even if the best fit parameters do not agree within errors with the fit to the \iras\ sample. This is because we find overlap between mock, observable galaxies representing each of these redshift bins. See the posterior distributions of observed and mock samples in \autoref{fig:fit_params}. The detection limits of each sample are responsible for any apparent evolution in the correlation between IR luminosity and peak wavelength. \label{tab:best_fit_params}}
    \centering
    \begin{tabular}{P{0.3\columnwidth}P{0.3\columnwidth}P{0.3\columnwidth}P{0.3\columnwidth}P{0.3\columnwidth}P{0.3\columnwidth}}
        \hline\hline
        Sample & Redshift Range & $\lambda_{\rm t}$ & $\eta$ & $\sigma_{\lambda}$ & Number of Galaxies \\
        \hline
        \iras\ & $0.025 < z < 0.19$ & $92 \rpm 2$ & $-0.09 \rpm 0.01$ & $0.046 \rpm 0.003$ & 511 \\
        \hatl\ & $0.05 < z < 0.1$ & $96 \rpm 3$ & $-0.08 \rpm 0.07$ & $0.031 \rpm 0.003$ & 729 \\
        & $0.1 < z < 0.2$ & $95 \rpm 2$ & $-0.079 \rpm 0.010$ & $0.030 \rpm 0.002$ & 926 \\
        & $0.2 < z < 0.3$ & $92 \rpm 4$ & $-0.11 \rpm 0.04$ & $0.032 \rpm 0.007$ & 328 \\
        & $0.3 < z < 0.4$ & $96 \rpm 2$ & $-0.05 \rpm 0.03$ & $0.029 \rpm 0.007$ & 184 \\
        COSMOS & $0.15 < z < 0.5$ & $97\rpm3$ & $-0.074 \rpm 0.009$ & $0.052 \rpm 0.006$ & 650 \\
        & $0.5 < z < 1.0$ & $96 \rpm 1$ & $-0.09 \rpm 0.01$ & $0.059 \rpm 0.004$ & 743 \\
        & $1.0 < z < 1.5$ & $101 \rpm 2$ & $-0.15 \rpm 0.03$ & $0.056 \rpm 0.005$ & 359 \\
        & $1.5 < z < 2.0$ & $96\rpm 5$ & $-0.11 \rpm 0.04$ & $0.067 \rpm 0.006$ & 238 \\
        \hline\hline
    \end{tabular}
\end{table*}

The \iras\ sample serves as our low reshift anchor to which we compare the \hatl\ and COSMOS samples that extend further out to high redshift.
The left panel of \autoref{fig:IRAS_lirtd} shows the \lirlp\ correlation fit for the \iras\ sample.
The right panel of \autoref{fig:IRAS_lirtd} shows a histogram of the log of the peak wavelength measurements minus the log of the best fit model peak wavelengths, illustrating the scatter about the \lirlp\ relation.
The distribution is well fit by a Gaussian with a tail of galaxies at shorter peak wavelengths, or hotter dust temperatures.
Active galactic nuclei (AGN) are a possible extra source of heating for dust in galaxies. They are capable of heating torus dust to temperatures from 50--100\,K to as high as the dust sublimation temperature, $\sim$1000--1500\,K \citep[e.g.][]{Donley12d}.
The galaxy-integrated dust temperature may or may not be shifted to warmer temperatures because of AGN, depending on the emergent AGN luminosity.
To remove bias in our fits to the \lirlp\ correlation due to the presence of the hotter tail of sources in the \iras\ sample we iteratively sigma clip sources at $>$3\,$\sigma$, consisting of 27 out of 511 sources (5\%).
We find that approximately 26\% of these outliers (7 out of 27) contain known active galactic nuclei by cross referencing with the first bright quasi stellar object survey \citep{Gregg96x}, the thirteenth edition of the catalogue of quasars and active nuclei \citep{Veron-Cetty10j}, and the all sky catalog of WISE-selected active galactic nuclei (AGN) from \citet{Secrest15w}. Note that the fraction of cold (``inliers'') similarly identified as AGN in the same catalogs is significantly lower (7\%, or 35 of 484) than 26\%.
For the other samples we do not sigma clip, as we do not find a substantial population of outliers like those in the \iras\ sample. The reason they are present in the \iras\ sample is likely due to the 60\,\um\ selection. Figure \ref{fig:det_lims_median_z} shows how a 60\,\um\ selection is more sensitive to low luminosity warm outliers compared to the selection functions that drive the \hatl\ and COSMOS samples. Though AGN are present in our samples and have been shown to account for about 10\%, on average, of the total IR luminosities of their host galaxies \citep[e.g.][]{Iwasawa11a}, they do not seem to affect the global relationship measured here.
Note that if we were not to clip these outliers from our sample, the derivation of the $\Theta_{\iras}$ parameter set would still be consistent with our findings within measurement uncertainty; however, we find that this clipping provides the most appropriate fit by excluding sources known not to follow the bulk relationship drawn from a normal distribution in $\log(\lambda_{\rm peak})$ about a given IR luminosity.

The best fit parameters for the \iras\ sample, $\Theta_{\iras}$, are given in \autoref{tab:best_fit_params}.
In the following subsections we will use these values as our basis for comparison with our other samples to investigate redshift evolution.

\begin{figure*}
    \centering
    \includegraphics[width=0.49\textwidth]{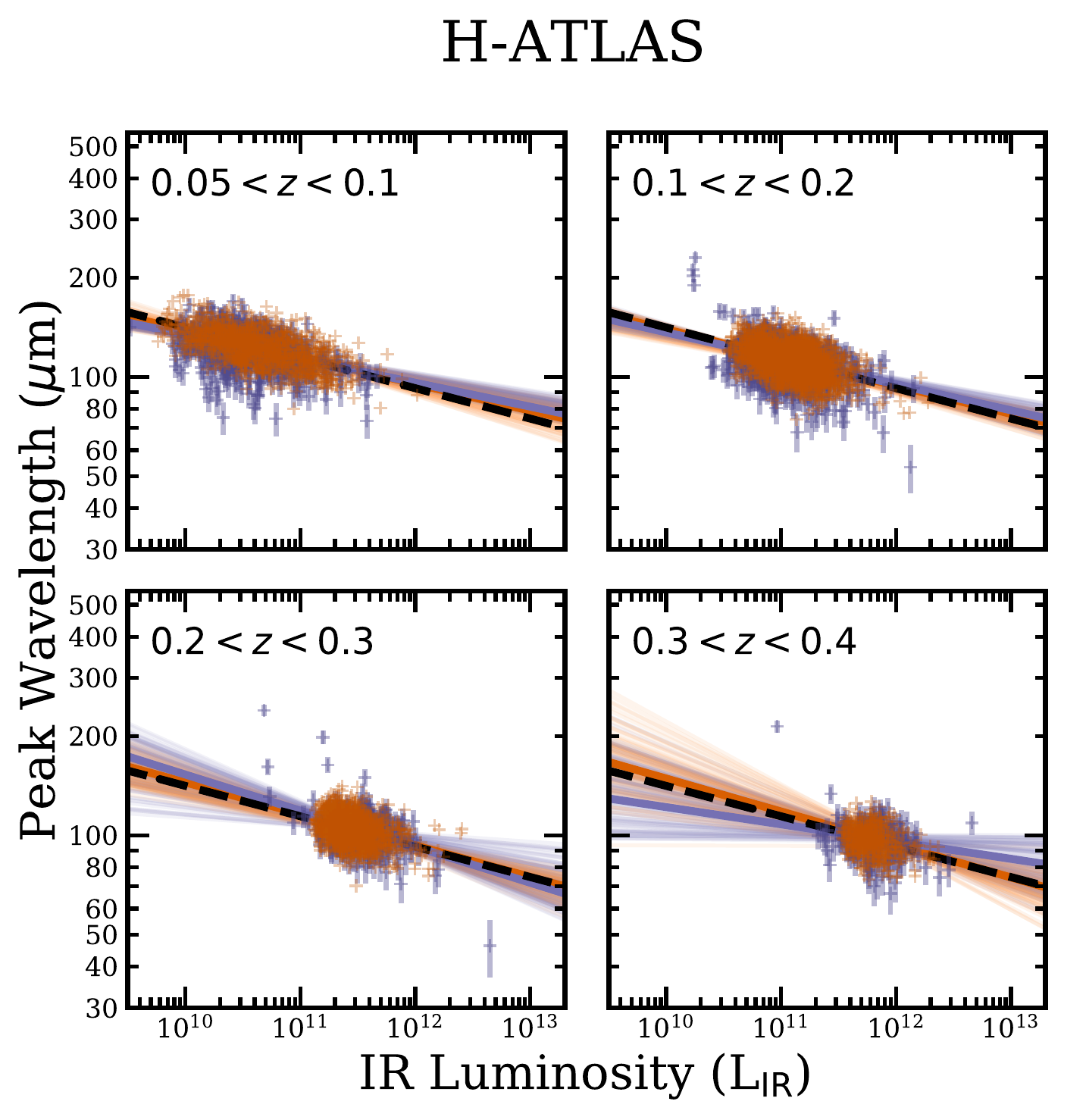}
    \includegraphics[width=0.49\textwidth]{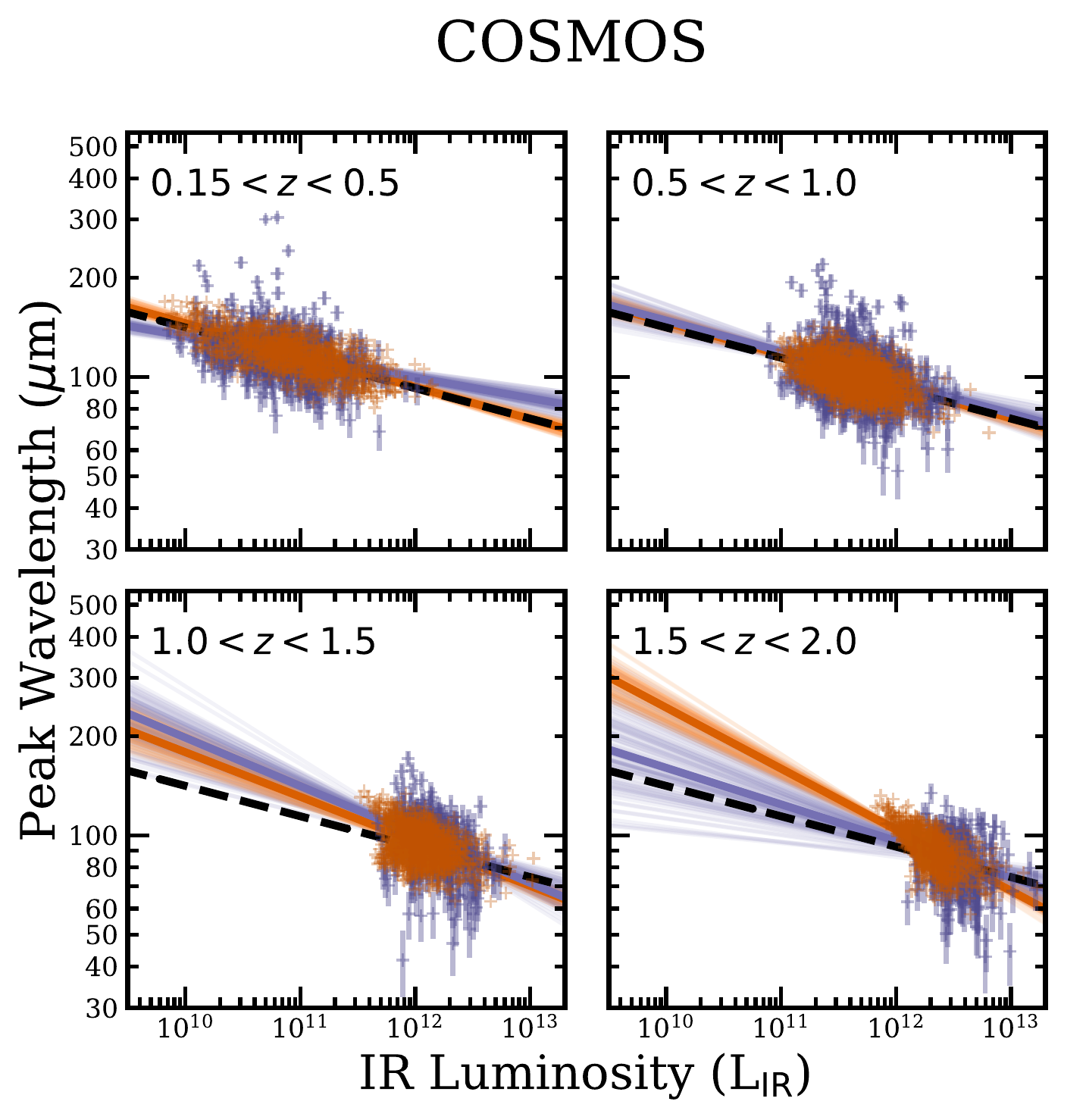}
    \caption{SED peak wavelength versus IR luminosity for the \hatl\ and COSMOS samples on the left and right, respectively, split into redshift bins. The purple points are individual observed galaxies and the orange contours depict the area containing 95, 68, and 34\% of mock galaxies. The light orange and purple lines indicate 100 randomly chosen MCMC chains, and the dark orange and purple lines show the best fit to the \lirlp\ correlation in the same color coding as above. The black dashed line depicts the best fit to the \iras\ sample. In all redshift bins, the best fits to the observed and mock data overlap, within errors with each other (see \autoref{fig:fit_params}), though not necessarily with the best fit to the \iras\ sample. This discrepancy is caused by the samples' detection limits causing some galaxies to be systematically undetectable. This highlights the importance for correcting for detection limits while interpreting galaxies' dust temperatures.}
    \label{fig:lirlp_hatlas}
\end{figure*}

%%%%%%%%%%%%%%%%%%%%%%%%%%%%%%%%%%%%%%%%%%%%%%%%%%%%%%%%%%%%%%%%%%%%
\subsection{Generating Mock Galaxies}\label{sec:model_galaxies}

Beyond the nearby universe, the detection limits of far-infrared/submm facilities can, depending on the wavelength of observation, be biased with respect to the galaxies' dust temperatures, rendering either very cold or very warm SEDs undetectable.
This is already demonstrated in our analysis by the presence of likely AGN-heated hot sources in the \iras\ sample (driven by the 60\,\um\ selection) not present in the \hatl\ and COSMOS samples.
If one tries to fit the \lirlp\ correlation for sub-samples of galaxies selected at different wavelengths without first accounting for these relative selection biases the measured fit parameters $\Theta$ may themselves be biased.
Such a bias is a function of sample selection wavelength, detection limit, and redshift.
\autoref{fig:det_lims_median_z} shows the detection limits at the median redshift of each sample plotted with contours of detected galaxies across all redshifts in each sample.
For galaxies to be detected in a given sample they need to have higher IR luminosities than indicated by the thick colored lines which denote the flux density selection process outlined in \autoref{tab:selections}.
For example, to be detected in the COSMOS sample, galaxies must lie to the right of at least two curves in the rightmost panel of \autoref{fig:det_lims_median_z}: each curve represents the detection limit in each of the five bands used to define the catalog.
For the \hatl\ sample, the inclusion of the 100\,\um\ detection requirement acts like a cut in IR luminosity; it is insensitive to dust temperature. We find this additional selection greatly improves our constraints on the measurement of peak wavelengths of galaxies in this sample.

To determine the degree to which measuring $\Theta$ at any given redshift is biased by selection functions we generate mock galaxies to represent each dataset in this paper. These mock galaxies are modeled such that they follow the local \iras\ sample derived \lirlp\ correlation. In other words, we assert in the mock samples that the \lirlp\ correlation does not evolve with redshift.
These mock galaxies are generated using the $\Theta$ parameter set fit to the \iras\ sample, and are therefore used to test the null hypothesis, i.e. that there is no intrinsic redshift evolution in $\Theta$ parameters.

In order to generate representative galaxies for each sample we use the infrared luminosity function from \citet{Zavala21f} based on the \citet{Casey18a} model to calculate the expected number of galaxies of each IR luminosity that exist in the given survey solid angle. Note that these IR luminosity functions are consistent with many other IR luminosity functions in the literature at $z\lesssim 2$ \citep[e.g.][]{Magnelli09a, Casey12b, Casey12c, Magnelli13a, Gruppioni13a}.
We randomly assign IR SEDs to the mock galaxies by sampling a normal distribution in log(\lp) with width $\sigma_{\lambda}$ around the best fit \lirlp\ correlation and with $\alpha$ and $\beta$ values drawn from the posterior distributions of the \iras\ sample.
Photometry is then generated for the mock galaxies at the wavelengths where observations exist for each sample with appropriate instrument and confusion noise added.
We then remove mock galaxies that fall below the detection limits of the survey as defined by the thresholds in \autoref{tab:selections} in each redshift bin.
Thus, we generate mock versions of each sample in both survey volume and selection biases as they would exist if the null hypothesis were true.
Then, if the measured $\Theta$ from the mock samples are statistically consistent with the $\Theta$ measured from the observed samples in a given redshift bin then that redshift bin is consistent with the null hypothesis, i.e. that there is no redshift evolution in the \lirlp\ relationship since $z\sim0$.

We test for selection bias in the \iras\ sample in this same manner, though by constuction we might expect agreement, given that mock galaxies are drawn from the derived parameter set $\Theta$ fit to the \iras\ sample.
This would not be the case if there was a strong selection bias, however.
We find that the fits to the mock \iras\ sample are consistent within errors with the measurements from the observed sample. 
This suggests that the detection limits for the \iras\ sample are not presenting a strong bias for our measurement of $\Theta$ in the nearby universe.
Additionally, as is detailed in $\S$\ref{subsec:iras_sample}, we find no offset between the inputs and best fit values of mock galaxies that we generate in a grid covering a larger range of \lirlp\ space than exists in the \iras\ data. Together, these two tests suggest that there is not a strong bias for our measurement of $\Theta$ in the nearby universe.

\subsection{Is There Evidence for Redshift Evolution in \lirlp?}
With mock samples that match our observed \hatl\ and COSMOS samples in hand, we investigate whether or not there is evidence of redshift evolution in the \lirlp\ relation. To do this, we divide both mock and observed samples into discrete redshift bins for which we measure $\Theta$.
We choose bin widths that provide an adequately large dynamic range in the subsamples' \lir\ ($\gtrsim$1\,dex) and at least a few hundred galaxies per bin. Both of which are needed to accurately constrain the parameters of $\Theta$ after accounting for the added uncertainty due to selection biases.
For the \hatl\ sample we use redshift bins of width $\Delta z = 0.1$ between $0.05 < z < 0.4$ with the lowest redshift bin truncated to $0.05 < z < 0.1$.
For the COSMOS sample we use bins with width $\Delta z = 0.5$ ranging between $0.15 < z < 2$, again with a truncated lowest redshift bin with $\Delta z = 0.35$.

Figure \ref{fig:lirlp_hatlas} shows the \hatl\ and COSMOS samples in the \lirlp\ plane in purple along side mock galaxies made to mimic them in orange.
We find that for the most part the mock galaxies qualitatively resemble the distribution of real galaxies in both the \hatl\ and COSMOS samples. The black dashed line shows the best fit to the \iras\ sample, and the orange and purple lines show 100 random samples from the MCMC fits to real and mock galaxies, respectively. Because the mock galaxies are drawn from a distribution mirroring the nearby \iras\ sample, we test for redshift evolution by comparing the best-fit parameter sets of the real samples to the mock galaxy samples. If the two fits are statistically consistent within a 95\% confidence interval (i.e. 2$\sigma$) that would constitute direct observational evidence that the \lirlp\ relation does not evolve.

\begin{figure*}
    \centering
    \includegraphics[width=0.39\textwidth]{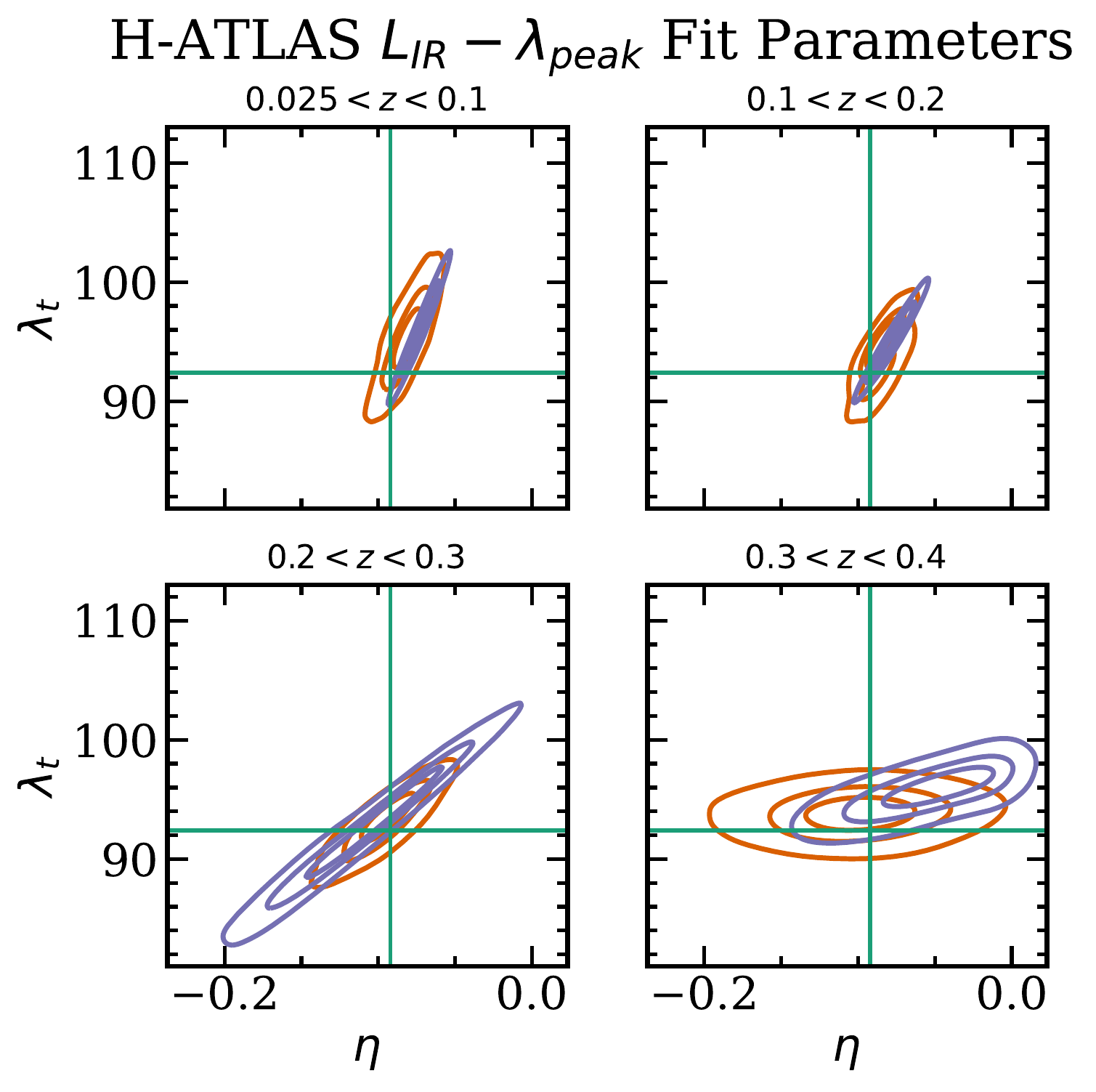}
    \includegraphics[width=0.39\textwidth]{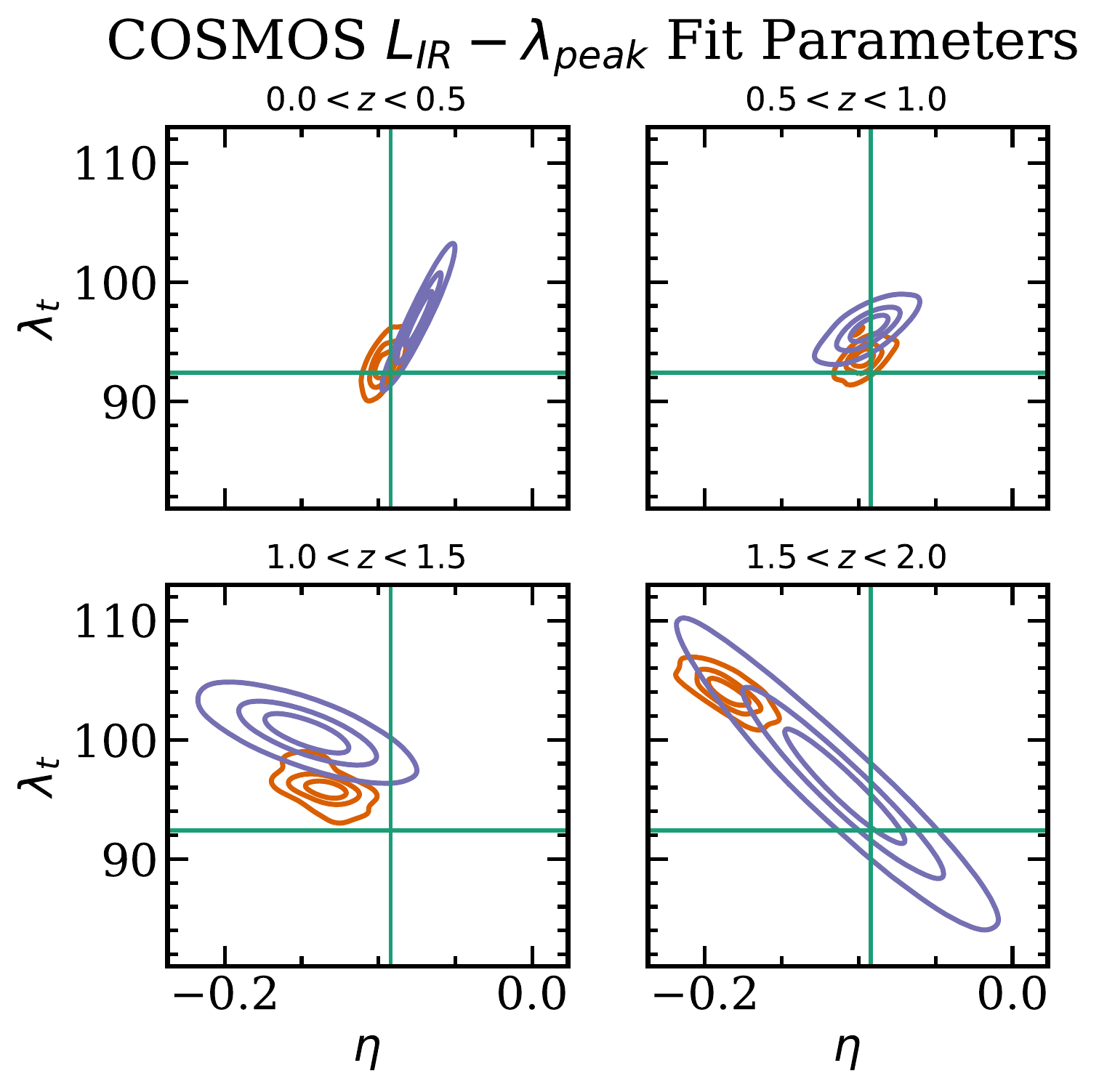}
    \includegraphics[width=0.157\textwidth]{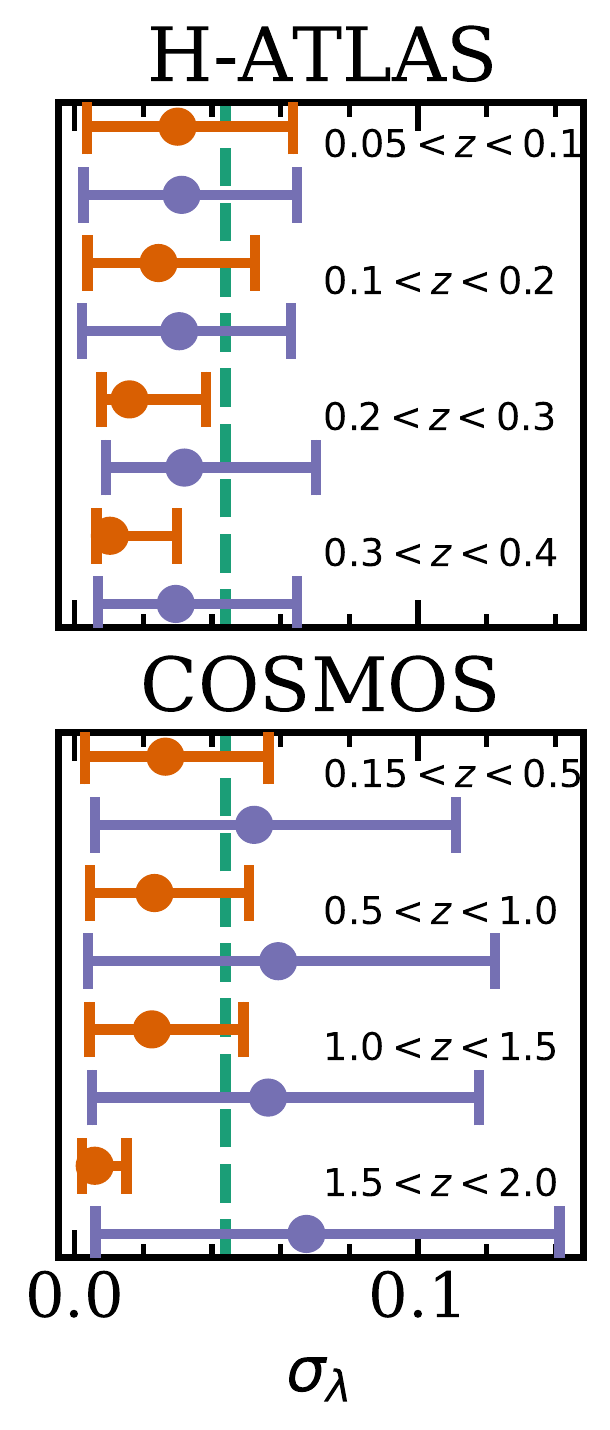}
    
    \caption{The four panels on the left show posterior distributions from MCMC fits to the \lirlp\ correlation for the \hatl\ sample in each redshift bin. The four panels to the right of those show the same for the COSMOS sample, and the two rightmost panels show the \lirlp\ distribution widths, $\sigma_{\lambda}$ in each redshift bin. Observed galaxies are shown in purple and mock galaxies are shown in orange.
    The green lines in all figures show the best fit parameters from the fiducial \iras\ sample with best fit parameters, $\log\lambda_{\rm peak} = \log (92) -0.09(\log{\rm L_{\rm IR}} -12)$. The best fit width is $\sigma_{\lambda} = 0.046\rpm0.002$. Note that measurement of correlation parameters for our data may be discrepant with the anchor parameters calculated from the \iras\ sample and still be consistent with no evolution if the fitted parameters are consistent with mock sample parameters at the same redshift and same filter set. This is because the mock samples are generated assuming no evolution; deviation from the input parameters is indicative of selection bias.}
    \label{fig:fit_params}
\end{figure*}

\autoref{fig:fit_params} compares the best-fit parameter set $\Theta=(\lambda_{\rm t}, \eta, \sigma_{\lambda})$ from \autoref{eq:lirlp_corr} between real and mock galaxies in the \hatl\ and COSMOS samples.
Because we find that variations in $\Theta$ associated with generating a volume limited set of mock galaxies are larger than the uncertainties derived when fitting any one set of mock galaxies ($\sim$2$\times$ higher than uncertainties) we simulate each redshift subsample for \hatl\ and COSMOS 100 times using input $\Theta$ that are drawn from the posterior distribution of the \iras\ sample, $\Theta_{\iras}$.
We then measure $\Theta$ in each trial to generate a distribution of measured $\Theta$ for each subsample.
Similarly, we also fit 100 bootstrap iterations of the real data for each subsample to generate realistic joint probability distributions of $\Theta$.

To statistically assess the agreement or disagreement of our measured parameter set $\Theta$ for a given sample and redshift bin, we compute the ratio of slopes and intercepts between mock and real samples, $\eta_{\rm mock}/\eta_{\rm real}$ and $\lambda_{\rm t,\,mock} / \lambda_{\rm t,\,real}$. If mock and real galaxies are statistically consistent, these ratios will be approximately equal to one. We find that indeed, $\eta_{\rm mock}/\eta_{\rm real}$ and $\lambda_{\rm t,\,mock} / \lambda_{\rm t,\,real}$ are consistent with 1 within the 95\% confidence interval (2$\sigma$) for all subsamples and redshifts. In other words, we rule out the possibility of an evolving \lirlp\ relationship at a confidence of $>$99.9997\% (or 4.6$\sigma$ significance), which we calculate by integrating the two-dimensional probability density distribution functions in derived parameters $\eta$ and $\lambda_{\rm t}$ at values that would be $>$3$\sigma$ offset from those measured in $\Theta_{\iras}$. Thus we conclude for all subsamples the data are consistent with no redshift evolution in the correlation between the IR luminosities of galaxies and the peak wavelength of their dust emission between $0 < z < 2$.

The rightmost panels of \autoref{fig:fit_params} show the width of the \lirlp\ distribution in each redshift bin for all samples. 
The best fit to the \iras\ is shown as a green dashed line, the scatter of observed galaxies about the \lirlp\ correlation, $\sigma_{\lambda}$, is shown in purple and the scatter of the distributions of mock galaxies is shown in orange. The measured $\sigma_{\lambda}$ of the real galaxies are consistent with the measurements both from the \iras\ sample and the mock samples for each redshift bin.

%%%%%%%%%%%%%%%%%%%%%%%%%%%%%%%%%%%%%%%%%%%%%%%%%%%%%%%%%%%%%%%%%%%%%%%%%%%%%%%%%%
\section{Discussion}\label{sec:discussion}

We have shown that the correlation between the peak wavelengths of dust emission and the IR luminosities of galaxies, i.e. the \lirlp\ relation, does not evolve between $0 < z < 2$.
We have done so by demonstrating that galaxy samples out to $z\sim2$ are fully consistent with the local relation when their selection wavelengths and depths are properly accounted for, in both sample selection and SED fitting. Similarly, we find that the scatter about the \lirlp\ relation, $\sigma_{\lambda}$, does not evolve out to $z\sim2$ within measurement uncertainty.

\subsection{Physical Interpretation of the \lirlp\ Relationship}
We interpret our results to mean that IR-luminous galaxies at high-redshifts are not dramatically different in size or geometry than those in the local universe, assuming the dust opacity does not significantly vary between sources or does not evolve with redshift.
The $z\sim0$ relation we measure in \lirlp\ from the \iras\ sample exhibits only a subtle IR--luminosity dependency on \lp\ as evidenced by measured slope of $\eta=-0.09\rpm0.01$. For context, over four decades of luminosity, from 10$^9$\,\lsun\ to 10$^{13}$\,\lsun, the mean \lp\ measured is only expected to change by a factor of $2.3 \rpm 0.3$. The intrinsic scatter about the relation, measured via the parameter $\sigma_{\lambda}$, is of order the same factor ($\times$1.1). The subtlety of the relationship and large intrinsic scatter can be interpreted with some understanding of galaxies' dust emitting sizes and source geometries.

The size and geometry of a dust emitting region relate directly to the emergent IR luminosity and peak wavelength as would be expected from the Stefan-Boltzmann Law \citep[e.g.][]{Hodge16a, Simpson17a}, where luminosity is proportional to dust temperature to the fourth power and to surface area (for an optically thick blackbody). A nice discussion of the applicability of the Stefan-Boltzmann Law to galaxies' dust emission is offered in \citet{Burnham21i}, who show that IR luminosity surface density is the most fundamental tracer of a galaxy's integrated dust temperature, or \lp\ \citep[see also][]{Chanial07e, Lutz16c}.
Using the equation relating the best fit surface density of IR luminosity to the peak wavelength of the dust SED from \citet{Burnham21i}, $\log\Sigma_{\rm IR} \approx 20.18 - 4.19 \log(\lambda_{\rm peak}/\mu{\rm m})$, we find that for a galaxy with \lir\ $= 10^{12}$\,\lsun\ the difference between galaxies at peak wavelengths
$\rpm$1$\sigma_{t}$ corresponds to a difference of only 0.2\,dex ($<$2$\times$) in the effective radius of the emitting dust.
A wide array of dust geometries in individual galaxies, from compact and clumpy to extended and diffuse, then accounts for the intrinsic scatter in the \lirlp\ relation at fixed \lir. This scatter translates to $\rpm$5\,K temperature differences at $\lambda_{\rm peak} = 100\,$\um, and an effective half-light radius difference of 0.4\,kpc for galaxies of typical radii of 0.8\,kpc, assuming the case of an optically thick blackbody.
Because galaxy sizes \citep[as measured by dust continuum, e.g.][]{Hodge16a, Simpson17a} only vary by a factor of a few and not orders of magnitude, the \lirlp\ relation itself has only shallow dependence on \lir. In other words, in our best-fit relation for the \iras\ sample, $\eta=-0.09\rpm0.01$ implies a size-luminosity relationship such that $R_{\rm eff} \propto {\rm L_{\rm IR}}^{0.31}$ \citep[consistent with $R_{\rm eff} \propto {\rm L_{\rm IR}}^{0.28\pm0.07}$ measured by][]{Fujimoto17g}.

Though this work does not directly measure the size of the dust emitting region in our samples, nor does it quantify the dust geometry through morphological analysis, our finding of consistent non-evolving SEDs in the \lirlp\ plane argues for unchanging dust sizes from $z\sim2$ to $z\sim0$. Similarly, it argues that the breadth of sizes in those galaxies also does not evolve. 
This naturally follows when considering the IR-luminous population ($>$10$^{11}$\,\lsun) is, on a whole, representative of fairly massive galaxies that have likely established most of their mass reservoirs --- whether in stars or gas supply that will be converted to stars --- and will not continue to grow substantially in size, whether they are at $z\sim2$ or $z\sim0$.

\subsection{Comparison with Literature Dust Temperatures: Evolution or No Evolution?}

Our finding that the \lirlp\ relationship does not evolve from $0<z<2$
is perhaps seen to be in conflict with other works in the literature
that claim galaxies' temperatures evolve with redshift.
One set of claims argues that higher redshift galaxies have hotter dust
temperatures compared with galaxies at $z\sim0$. This is based on observations of galaxies between $1 \lesssim z \lesssim 5$ \citep[e.g.][]{Magdis12a, Magnelli14a, Bethermin15a, Faisst17b, Schreiber18a, Faisst20a}, and theoretical modeling, primarily at $z\gtrsim5$ \citep[e.g.][]{De-Rossi18w, Ma19a, Liang19a, Arata19a, Sommovigo20a}.
These higher temperatures are attributed to higher specific star formation rates, higher star formation surface densities, or lower dust abundances relative to $z\sim0$ samples.

Another set of claims argues that dust temperatures are \textit{colder} at high redshift than at $z\sim0$ \citep[e.g.][]{Chapman02g, Pope06a, Symeonidis09y, Symeonidis13a, Hwang10a, Kirkpatrick12a, Kirkpatrick17a, Magnelli14a}.
This is attributed to galaxies having higher dust masses \citep[e.g.][]{Kirkpatrick17a}, higher dust opacities or emissivities, or more extended spatial distributions of dust at higher redshift \citep[e.g.][]{Symeonidis09y, Elbaz11a, Rujopakarn13o} as compared with $z\sim0$.

A third set of claims argues that the dust temperatures of galaxies do not evolve with redshift. These works find no evidence of redshift evolution in \lirlp\ space for their samples \citep[e.g.][]{Casey18a, Dudzeviciute20a, Reuter20a}, which we also find.
How is it possible that there are such seemingly contradictory conclusions in the literature?

\begin{figure}
    \centering
    \includegraphics[width=0.99\columnwidth]{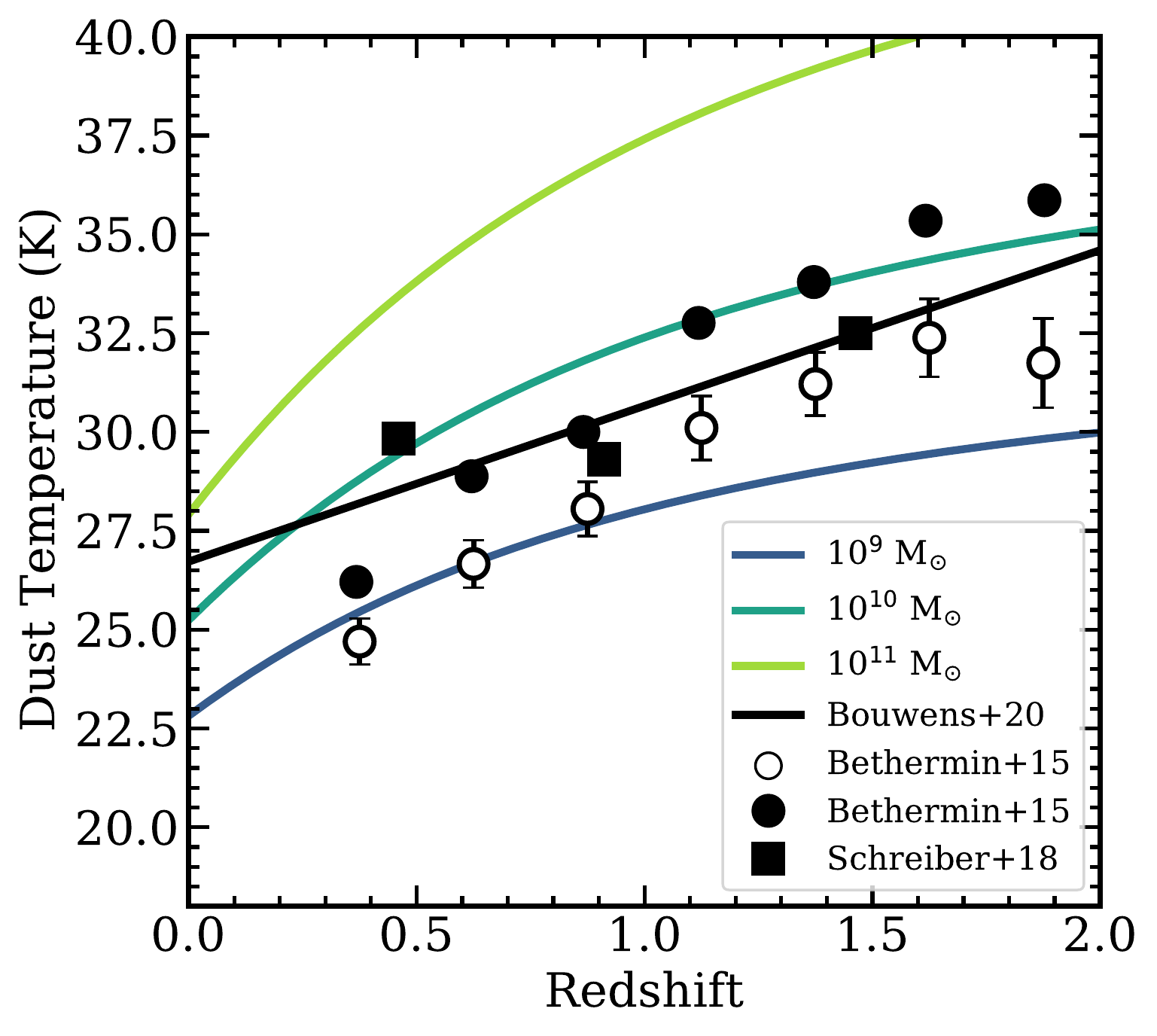}
    \caption{Dust temperatures as reported for stellar mass-selected samples of galaxies from the literature as a function of redshift (black points). Colored curves are generated assuming a fixed \lirlp\ relation (as shown in \autoref{fig:IRAS_lirtd}) joined with an evolving SFR--\mstar{} relation from \citet{Speagle14a}. Each curve represents a different fixed stellar mass as indicated in the legend. Note that the \citet{Bethermin15a} black circles, indicative of average SEDs for galaxies between 10$^{10-11}$\,\msol{}, are drawn from the work of \citet{Bouwens20a}. We have refit the original photometry from \citet{Bethermin15a} using the SED fitting procedure herein and the results are plotted as open circles. 
    We restrict the plotted data to $0 < z < 2$ because that is the redshift range over which we constrain galaxies' SEDs.
    Each sample has adopted different SED fitting techniques, making the temperatures not directly comparable with one another, though within a sample the monotonic increase in temperature is still observed. We find the observed increases consistent with the expected trend for galaxies of fixed stellar mass and a non-evolving \lirlp\ relation.
    }
    \label{fig:bouwens}
\end{figure}

Our results are, indeed, consistent with more than just the third set of these claims.
Observational works such as those by \citet{Magnelli14a}, \citet{Bethermin15a}, and \citet{Schreiber18a} find that dust temperatures increase between low and high redshift by studying samples of galaxies at fixed stellar mass on the main sequence. The star formation rates of galaxies on the main sequence increase as redshift increases \citep[e.g.][]{Speagle14a} which naturally results in an increase in IR luminosity. 
\autoref{fig:bouwens} highlights how a non-evolving \lirlp\ relation then translates to a perceived evolution in dust temperatures when drawing from main sequence galaxies at fixed stellar mass. 
At high redshift, the SFR (thus \lir) at fixed stellar mass is higher \citep[e.g.][]{Speagle14a} and galaxies with higher IR luminosities have hotter SEDs \citep[e.g.][]{Casey18a, Burnham21i}.
The tracks of fixed
stellar mass in this figure are derived from our \lirlp\ relationship, where
\lir\ is converted to a star-formation rate \citep{Kennicutt12a}
then to stellar mass (using the \citeauthor{Speagle14a} \citeyear{Speagle14a} relation).  We
then convert \lp\ to dust temperature using a simple optically-thin
modified blackbody, as in \citet{Bouwens20a}. We see
remarkable agreement between our projected \lirlp\ curves at fixed
stellar masses and measurements from literature works claiming
measurement of hotter SEDs at higher redshifts \citep[e.g.][]{Magnelli14a, Bethermin15a, Schreiber18a}.  While
the conversion between \lp\ and dust temperature is reliant on an
understanding of the underlying opacity of dust, there is roughly linear scaling between the dust
temperatures estimated using the optically thin assumption and other
opacity models. So while the absolute dust
temperatures may not be known, the trend toward ``hotter'' SEDs as
perceived in \autoref{fig:bouwens} should be relatively robust.
An illustration of this can be seen with the \citet{Bethermin15a}
sample. Their calculated dust temperatures (black circles) do not agree with those
that we calculate for their sample (having refit them using MCIRSED; open circles), though both show monotonic increase in temperature with
redshift. This offset is due to differences in the model assumed to translate \lp\ to \td\ between our two works. We attribute the increase in temperature with redshift
solely to the increasing SFR of galaxies on the main sequence, in
line with expectations, given a non-evolving \lirlp\ relation.

\begin{figure}
    \centering
    \includegraphics[width=0.99\columnwidth]{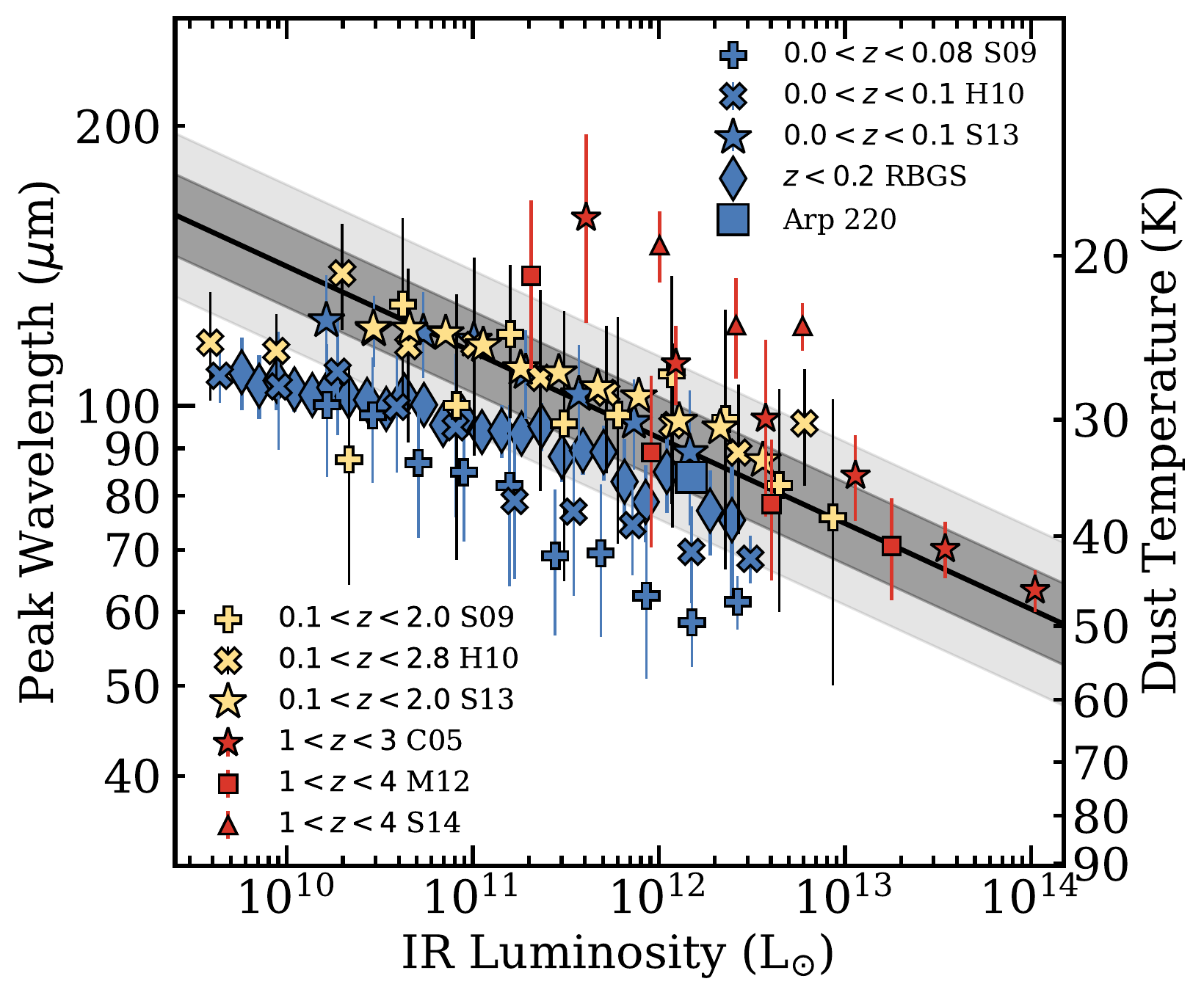}
    \caption{The \lirlp\ correlation, presented in averaged \lir\ bins, as explored by different datasets in the literature across a range of redshifts. Blue data points represent data at low redshift ($z \lesssim 0.2$), cream colored points span a range of redshifts from low to high within each sample (from $z \sim 0.2$ to $z\sim2$), and red points show samples at high redshift ($z\gtrsim2$). We plot our best fit correlation in black with $\rpm$1$\sigma$ in dark grey with $\rpm$2$\sigma$ in light grey. These datasets may be seen as having an apparent redshift evolution toward colder dust temperatures at higher redshift (at fixed \lir), however we find this is due to biases in SED fitting for local samples whose data, in most cases, is practically limited to data blue-ward of the dust peak (see \autoref{fig:coverage_bias}). Works plotted: \citet{Sanders03c, Chapman03b, Chapman05a, Symeonidis09y, Symeonidis13a, Hwang10a, Magnelli12a, Casey14a, Swinbank14a}.
    }
    \label{fig:lit_lirlp}
\end{figure}

In contrast to works claiming that higher-redshift galaxies have hotter temperatures, there are also a number of literature works that show higher redshift galaxies may have \textit{cooler} SEDs at fixed \lir\ \citep[e.g.][]{Chapman02g, Pope06a, Symeonidis09y, Symeonidis13a, Hwang10a, Kirkpatrick12a, Kirkpatrick17a,Magnelli14a}.
\autoref{fig:lit_lirlp} shows \lirlp\ correlation as reported from a handful of different works in the literature, color coded roughly by redshift. Blue points represent samples at low redshift, red points represent samples of high redshift galaxies, and cream points are representative of samples at intermediate redshifts. We have converted dust temperatures to
$\lambda_{\rm peak}$ using the fixed values of $\beta$ and $\lambda_0$ from each sample, where provided by the authors, otherwise we use Wien's Law to convert, as was done in the case of \citet{Chapman05a}, \citet{Magnelli12a}, and \citet{Hwang10a}. We plot these works because we are interested in the general trend with redshift, not the absolute scaling of dust temperature. From what is presented here
from the literature, there is a clear offset between the lowest
redshift samples and the high redshift samples, with some of the intermediate redshift
samples aligning with the low redshift relation and some with the high
redshift relation.

\begin{figure}
    \centering
    \includegraphics[width=0.99\columnwidth]{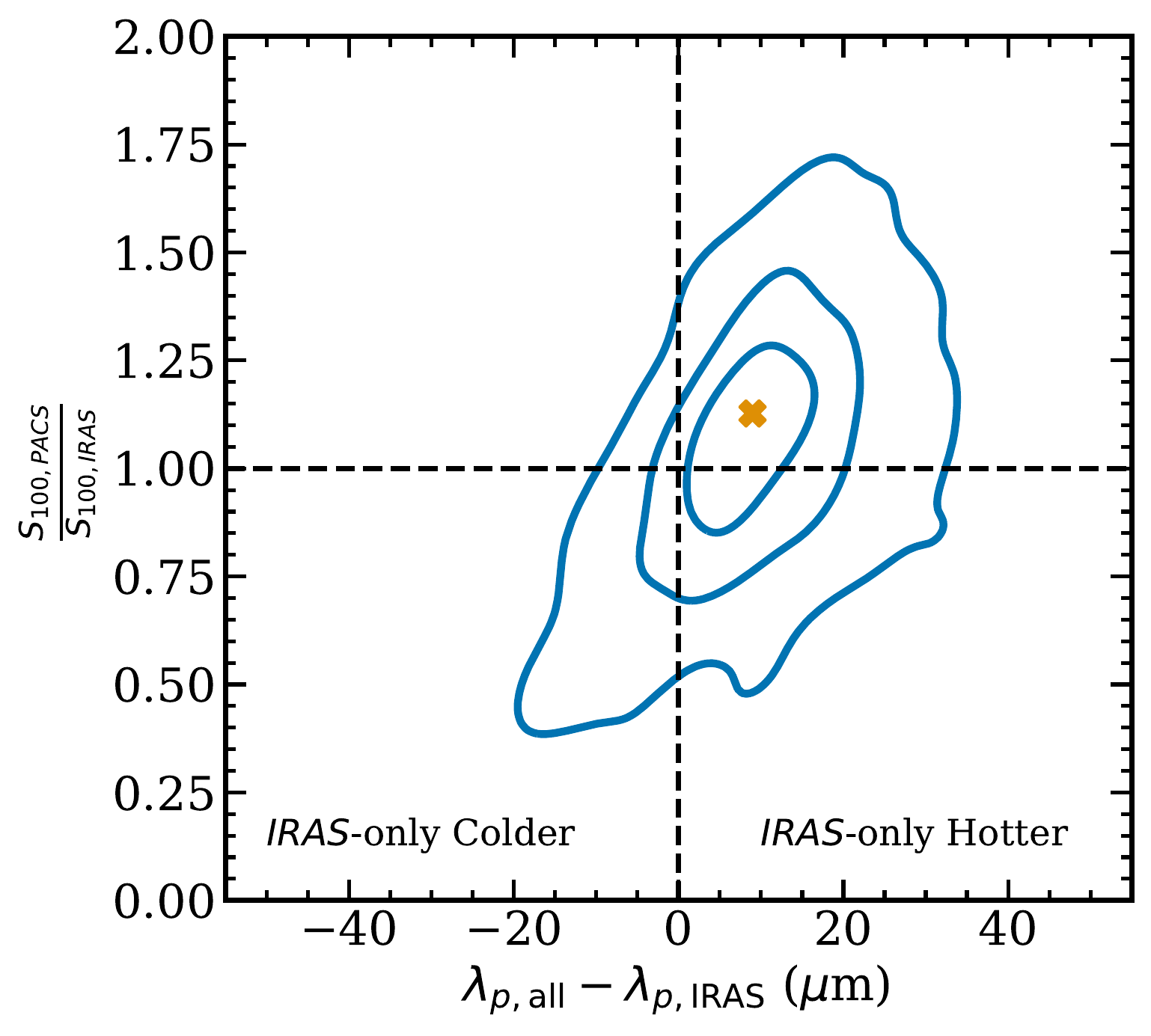}
    \caption{Contours showing the ratio of PACS flux densities to \iras\ flux densities at 100\,\um\ from our \iras\ sample versus the difference between peak wavelength as measured when fitting SEDs using all wavelengths in our \iras\ sample minus peak wavelength as measured when fitting only \iras\ wavelengths with the SED fitting code, CMCIRSED from \citet{Casey12a}.
    The yellow cross shows the median offset from $S_{100,\,PACS} = S_{100,\,IRAS}$ and $\lambda_{p,\,{\rm all}} = \lambda_{p,\,{IRAS}}$ (The black dashed lines). 
    % The median ratio between PACS and \iras\ flux densities at 100\,\um is 1.1$\rpm$0.3 and the offset in peak wavelength is 9\,\um.
    This offset is responsible for some of the claims in the literature that local galaxies have hotter dust temperatures than galaxies at the same IR luminosity at higher redshift.
    There are two effects at play here: 1) When fitting observations solely on the Wien side of the SED the fitted dust temperature is sensitive to the flux density of the longest wavelength of observation. If the SED truly peaks at longer wavelengths than the longest observed wavelength the fitted dust temperature will be warmer than the true dust temperature.
    2) The systematic offset between PACS and \iras\ observations at 100\,\um\ further causes fits to only \iras\ data to be warmer than those fit to \iras\ and \hersch\ data, as \iras\ is systematically lower, thus suggestive of a shorter \lp.}
    \label{fig:coverage_bias}
\end{figure}

In this paper we find no such evolution.  We attribute the difference
in conclusions to the manner of constraining these galaxies' SEDs to
measure \lp. Some of these works, in an effort to compare to the most
complete local samples of galaxies, used the \iras\ RBGS without
requiring submillimeter photometry just redward of the peak (e.g. \citeauthor*{Casey14a} \citeyear{Casey14a}). Indeed,
such observations did not exist for large numbers of galaxies until the launch of {\it Herschel}.
\autoref{fig:coverage_bias} shows that when
fitting SEDs to \iras\ data alone from local samples the
measurement of \lp\ is biased toward warmer dust temperatures than when fitting to data covering the full range of the dust SED.
Without data in the
160--500\um\ range, intrinsically colder SEDs (with
\lp\,$\simgt$\,100\um) become untethered, with very poor
\lp\ constraints.
On average, we find that {\it IRAS} photometry
alone will systematically underestimate \lp\ for such systems,
i.e. skewing fits toward hotter dust SEDs than they may intrinsically have.
Furthermore, we find a systematic offset between \iras\ 100\,\um\ measurements and \hersch\ PACS observations at 100\,\um. The median ratio of flux at 100\,\um\ from PACS to that from \iras\ for our \iras\ sample is 1.13$\rpm$0.01), shown in \autoref{fig:coverage_bias}. This systematic bias for the sample's 100\,\um\ flux densities results in warmer fitted dust temperatures when SEDs are fit to \iras\ data exclusively rather than the full suite of \iras, \wise, and \hersch\ described in $\S$\ref{subsec:iras_sample}.
With the introduction of {\it Herschel}'s photometric
constraints on the Rayleigh-Jeans tail, such biases are eliminated.  The stringent requirement we place on our local
universe {\it IRAS} sample, to have {\it Herschel}-SPIRE constraints
from H-ATLAS, this translates to colder net SEDs in the \lirlp\ relation
than had been calculated previously.  We have effectively removed the
bias whereby colder systems had been fit to SEDs with warmer
temperatures for lack of long wavelength data.

One last comparison we offer is a comparable fit to the Arp\,220 system, one of the best known ULIRGs in the nearby universe.
Literature works often describe Arp\,220 as potentially having a hotter dust SED than the average ULIRG at high redshift \citep[$\lambda_{\rm peak, Arp 220} \sim 60$\,\um\ versus $\lambda_{\rm peak,SMG}\sim70$--80\,\um, e.g.][]{Dudzeviciute20a}. However, when we refit Arp\,220's well sampled IR/mm SED \citep[e.g.][]{Eales89t, Rigopoulou96q, Sargsyan11b, Brown14m} using MCIRSED we find that, in fact, it is well within the expected scatter for galaxies of its luminosity at any redshift $0 < z < 2$ (see \autoref{fig:lit_lirlp}).

\subsection{Evolution of temperatures at $z>2$?}

We do not analyze galaxies beyond $z\sim2$ because \hersch\ SPIRE observations beyond that redshift are limited to very bright, rare sources (${\rm L_{\rm IR}} \gtrsim 10^{13}$\,\lsun).
Extending the analysis to higher redshift would require sufficient dynamic range in \lir\ with full SED sampling.
Sources in our samples at $z > 2$ do not provide enough dynamic range per redshift bin to perform this analysis and other literature samples either lack statistics or an unbiased sampling of SEDs.
We have demonstrated the importance of using observations that bracket the peak wavelength of the dust SED when trying to accurately measure \lp. 
Unfortunately, other samples do not bracket the peak wavelength of dust emission until $z\sim7$, when the observed-frame peak reaches $\sim$850\,\um, which is measureable from the ground to great depth thanks to ALMA.
Though there exist a number of works that fit dust temperatures to galaxies beyond $z > 2$, including SCUBA-2 and South Pole Telescope (SPT) samples \citep[e.g.][]{Dudzeviciute20a, Reuter20a}, these samples do not have enough galaxies per redshift bin to allow for the analysis presented in the present work to be performed.
Investigating whether there is evolution at $z > 2$ will require mJy sensitivity in the 250--500\,\um\ wavelength range, which does not currently exist (1\,mJy at e.g. 500\,\um\ corresponds to a galaxy of \lir\ = $8.4\times10^{10}$\,\lsun\ at $z=4$). Earth's atmosphere is opaque at those wavelengths whenever the atmosphere is not extremely dry. This constraint limits surveys to small areas and hence, small samples of galaxies.
One such survey, reaching a $\sim$1\,mJy depth at 450\,\um\ is STUDIES \citep{Wang17x}, which covers a 151\,arcmin region in the COSMOS field and detects too few sources to perform an analysis such as the one presented in this paper.
Future space-based FIR missions like the proposed Origins Space Telescope or the 2020 decadal recommendation of a FIR probe could provide these constraints for the $z > 2$ universe.
In the absence of direct constraints on galaxy SEDs down to $\sim$1\,mJy, stacking analyses can provide helpful insight but the different stacking techniques used in the literature in highly confused FIR maps can result in a broad range of predictions that are not especially constraining \citep[e.g.][]{Viero13a, Bethermin15a}.

\subsection{Recommendations for Fitting Dust SEDs}
The analysis presented in this paper has shown that disparate
techniques in IR/mm SED fitting and galaxy sample selection can
manifest in very different interpretations of galaxy
dust SEDs.  Our finding of broad consistency between SEDs from the local
Universe out to $z\sim2$ was shown only by accounting for both sample
selection wavelengths and by approaching SED fitting using a uniform
technique. Our SED fitting technique has been tuned to be broadly
applicable to most dusty galaxies in the literature that may lack
extensive photometric constraints in the IR/mm.

Based on work in this paper, we have assembled a list of ``best
practices'' in implementing IR/mm SED fitting for the community which
will facilitate easier cross-reference comparisons and eventual
measurement of broad physical trends in galaxy populations out to
higher redshifts.
These recommendations are:
\begin{enumerate}
\item In most cases, galaxies have a dearth of IR/mm data (3--5
  photometric constraints) and as a result, should not be fit to
  models with more free parameters than data. Though this is standard scientific practice, it is sometimes violated given the intrinsically complex nature of SED models and the relative dearth of data, especially in this wavelength regime. Most galaxy IR/mm SEDs
  are superpositions of modified blackbodies that may be approximated
  as a dominant cold blackbody joined with a mid-infrared powerlaw
  with no more than five free parameters.
\item The number of free parameters in IR/mm SED fitting should be
  adjusted according to existing data constraints.  Gaussian priors
  for parameters should be used where data are low signal-to-noise
  (i.e. SNR\,$<$\,5) and parameters should be fixed in the absence of data.  For
  example, the mid-infrared powerlaw slope (when using MCIRSED or other piecewise modified blackbody with infrared power law) should be set to $2.3\rpm0.5$ if fewer than two photometric constraints exist at
  rest-frame wavelengths $\lambda\simlt80$\,\um; similarly if fixing $\lambda_0=200$\,\um\ (as we have done here), $\beta$
  should be set to $1.96\rpm0.43$ if there are fewer than two
  photometric constraints longward of rest-frame $\simgt$250\,\um. These are the median parameter set values for our fitted \iras\ sample.
\item $\lambda_{0}$ should be fixed if there are no direct constraints
  on the dust column density via spatially resolved observations used to constrain dust mass surface density.
\item Dust temperatures cannot be directly constrained for the vast
  majority of galaxies that have poorly-sampled SEDs and no
  spatially-resolved dust continuum observations.  In lieu of dust
  temperatures, SED fitting should focus on measurement of \lp, the
  rest-frame peak wavelength of the SED in S$_\nu$.  Measurements of
  \lp, which serves as a proxy for dust temperature, are directly
  comparable between works, even when different SED modeling
  techniques are employed (assuming the dust opacities are roughly comparable). The same is not true of dust temperatures.
\end{enumerate}
We have made the MCIRSED code publicly available\footnote{{Publicly available at \href{https://github.com/pdrew32/mcirsed}{github.com/pdrew32/mcirsed}.}} 
to help facilitate the use of these
recommendations.  While this list should be regarded as a useful
guide, it requires custom attention, given the above recommendations.
Similarly, galaxies with especially thorough IR/mm coverage (e.g. like
some local (U)LIRGs, and high-redshift lensed objects
and quasars) may not be fit well by our modeling technique, in which
case more detailed radiative transfer models may be appropriate.  In
all instances, we recommend that community members first reflect on
their goals in deriving an IR/mm SED for a galaxy or sample of
galaxies before approaching the task of SED fitting, and use the
appropriate tool for the type of measurement that needs to be made.

\section{Summary}\label{sec:summary}
We have fit the dust SEDs of $\sim$4700 galaxies from different studies in the literature that span redshifts $0 < z < 2$ to investigate possible redshift evolution in galaxies' dust temperatures. This analysis is carried out within the context of the \lirlp\ plane, or between galaxies' IR luminosities and their rest-frame peak wavelengths of their dust SEDs, the latter of which is a model-independent\footnote{While \lp\ is model independent, the derivation of dust temperature from a given \lp\ is model-dependent and relies on some knowledge of a given galaxy's dust opacity.}.
We find no evolution in the \lirlp\ relation between $0 < z < 2$ which implies that, at fixed \lir, galaxies at $z\sim0$ are relatively similar to those at $z\sim2$. Specifically, we rule out the possibility that the slope and intercepts of the \lirlp\ relation evolve at $0 < z < 2$ with 99.9997\% confidence (corresponding to 4.6$\sigma$).
We find that the scatter about the \lirlp\ relation is also unchanged from $z\sim2$ to $z\sim 0$. Such scatter likely traces the diversity of dust sizes and dust geometries in IR-luminous galaxies.

We parameterize the \lirlp\ correlation as a Gaussian distributed about a line such that $\log(\lambda_{\rm peak}) = \log(\lambda_{\rm t}) + \eta [\log(\rm L_{\rm IR}) - 12] + N(\mu, \sigma_{\lambda})$ where $\lambda_t = 92\rpm2$\,\um, $\eta = -0.09\rpm0.01$, and N is a Gaussian distribution with $\mu \equiv 0$ and $\sigma_{\lambda} = 0.046 \rpm 0.003$.
This result is consistent with other works in the literature that have found that the dust temperature of galaxies does not evolve with redshift over the range $0 < z < 4$ \citep[e.g.][]{Casey18a, Dudzeviciute20a, Simpson19a, Reuter20a}.
This result is also consistent with seemingly conflicting claims in the literature which argue that dust temperature evolves with redshift and that galaxies are hotter at high redshift \citep[e.g.][]{Bethermin15a, Schreiber18a}.
This conflict is resolved when we consider that the \lirlp\ relation does not evolve but the SFR--\mstar{} relation does and the measured increasing temperatures are a function of samples with fixed stellar masses.
Our result also conflicts with claims in the literature which argue that dust SEDs are \textit{colder} at higher redshift than in the local universe \citep[e.g.][]{Pope06a, Symeonidis09y, Symeonidis13a}. The data was often limited in these samples to \iras\ observations only in the low redshift reference sample with which the high redshift samples are compared and we find that fitting SEDs to \iras\ data only can bias to warmer dust temperatures. In the present work we fit high quality data that spans the full range of the dust SED and find no such offset in temperature between local galaxies and those at higher redshifts.

Accounting for selection effects is crucial in the analysis of galaxies' IR/mm SEDs. We model selection bias by generating mock galaxies that are distributed in \lirlp\ space as they would be with the case of no evolution in the \lirlp\ correlation between $0 < z < 2$.
We then test for statistical deviation of the true data from the modeled mock samples and find that all the samples we have drawn from out to $z\sim2$ are consistent with the hypothesis that \lirlp\ does not evolve.

At present, the analysis performed in this paper cannot be extended to redshifts above $z\sim2$ because sensitive, high signal to noise observations are not available for large samples of galaxies in redshift bins which would allow for the study of redshift evolution.

We recommend that the number of free parameters used in dust SED fitting be adjusted according to the number of existing data constraints. Accurate measurement of the peak wavelength of dust emission depends on having high quality constraints on both sides of the SEDs true peak wavelength.

\section*{acknowledgements}
We thank Brendan Bowler, Lina Fernanda Gonz\'{a}lez Mart\'{i}nez, Milos Milosavljevic, Justin Spilker, Kre\u{s}imir Tisani\'{c}, and Jorge Zavala for help and useful discussions.
P.D. acknowledges financial support by NSF grants AST-1714528 and AST-1814034, and the University of Texas at Austin College of Natural Sciences.
This research has made use of the NASA/IPAC Infrared Science Archive, which is funded by the National Aeronautics and Space Administration and operated by the California Institute of Technology.
This research made use of Astropy, a community-developed core Python package for Astronomy \citep{Astropy13a}, and the python packages Matplotlib \citep{Matplotlib07a}, Numpy \citep{Numpy11a}, and Pandas \citep{McKinney10a}.
This research also made use of the NASA/IPAC Extragalactic Database (NED),
which is operated by the Jet Propulsion Laboratory, California Institute of Technology, under contract with the National Aeronautics and Space Administration.

\appendix
\section{Bayesian Modeling with PyMC3}\label{sec:bayesian_modeling}
\begin{figure*}
    \centering
    \includegraphics[width=0.99\columnwidth]{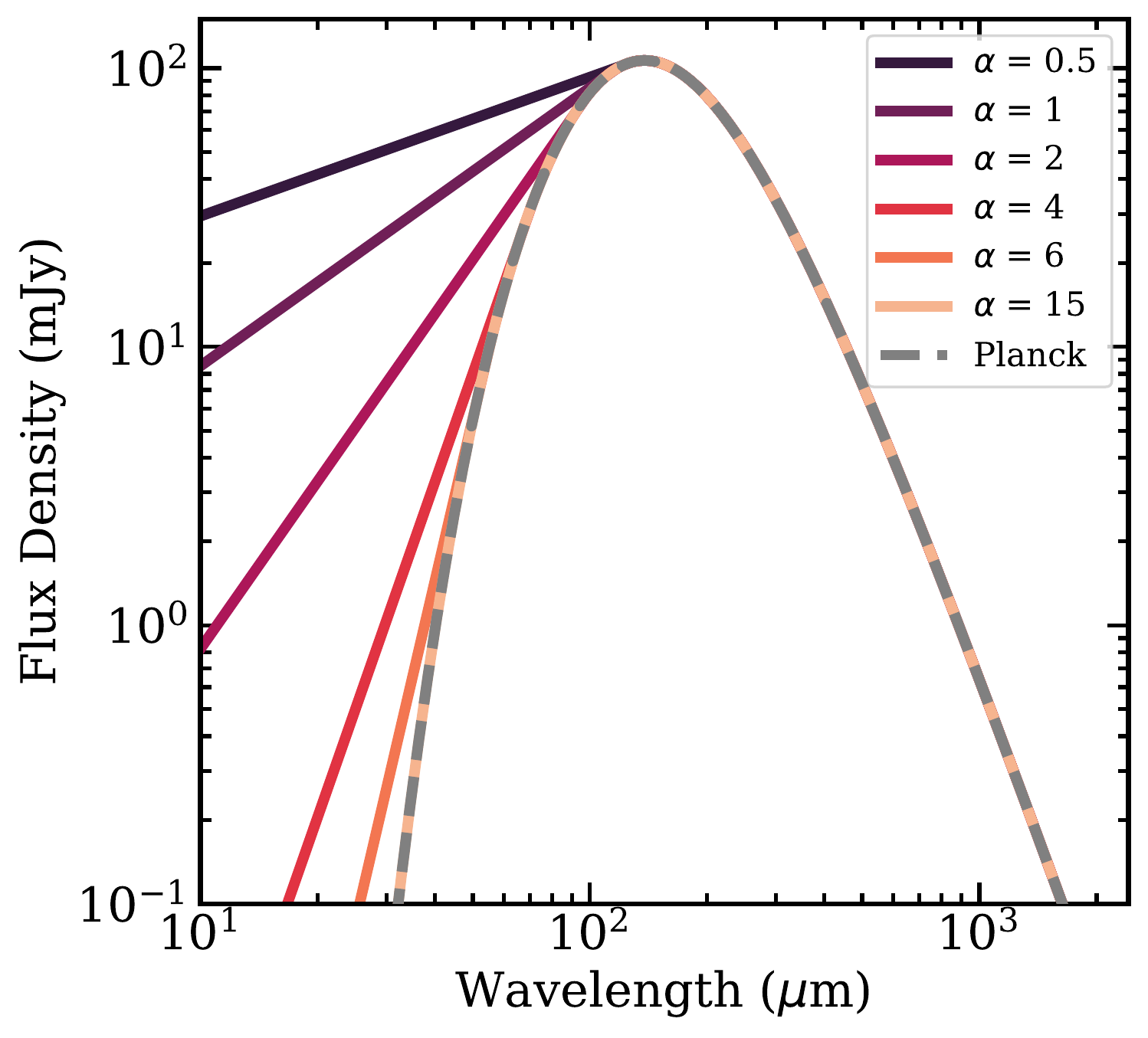}
    \includegraphics[width=0.99\columnwidth]{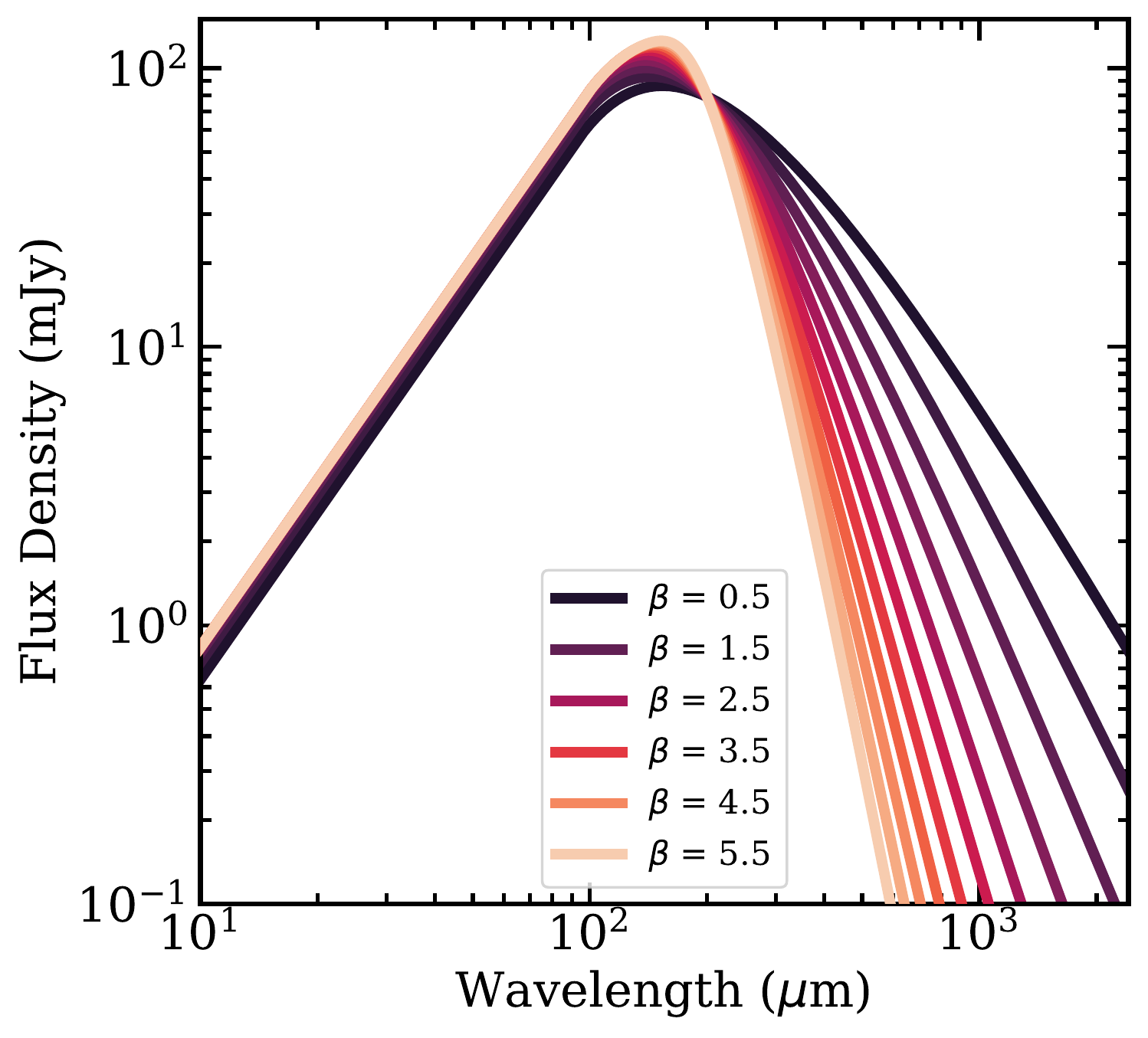}
    \caption{Left: Example SEDs with fixed dust temperatures, $\beta$, and $\lambda_0$ but varying $\alpha$ values ranging from 0.5 to 15. The $\alpha=15$ curve and the Planck function shown as a grey dashed line are indistinguishable because higher values of $\alpha$ asymptotically approach the Planck function at measurable flux densities. 
    Right: Example SEDs with fixed dust temperatures, $\alpha$, and $\lambda_0$ but varying $\beta$ values ranging from 0.5 (darkest) to 5.5 (lightest). Empirically, some of the highest observed dust emissivity values are $\beta = 3.7$ \citep{Kuan96h}.}
    \label{fig:different_alphas}
\end{figure*}
The model we employ to fit the flux densities of galaxies between 8--1000\,\um\ is:
\[S(\lambda) =
\begin{cases}
N_{pl}\lambda^{\alpha} & \text{: } \frac{\partial\log S}{\partial\log \lambda} > \alpha \\

\frac{N_{bb} \left(1-e^{-(\lambda_{0}/\lambda)^{\beta}}\right) \lambda^{-3}}{e^{hc/\lambda kT}-1} & \text{: } \frac{\partial\log S}{\partial\log \lambda} \leq \alpha,
\end{cases}\]
where $\lambda$ is the rest-frame wavelength, $\alpha$ is the mid-infrared power law slope, $N_{pl}$ and $N_{bb}$ are normalization constants, $\lambda_0$ is the wavelength where the dust opacity equals unity, $\beta$ is the dust emissivity index, $T$ is the luminosity-weighted characteristic dust temperature, $h$ is the planck constant, $k$ is Boltzmann constant, and $c$ is the speed of light.
$N_{pl}$ and $N_{bb}$ are tied to each other and to \lir. 
At short wavelengths, the SED is dominated by a power law with slope $\alpha$.
Physically, this shorter wavelength regime represents the warmer dust emission originating near star forming regions or active galactic nuclei whose dust mass is distributed with a power law distribution of temperatures \citep[e.g.][]{Kovacs10r}.
At long wavelengths, the SED is dominated by a modified Planck function with a self-absorption factor, $S(\lambda) \propto \left(1-e^{-\tau}\right)B_{\nu}(T)$,
with optical depth given by $\tau=(\lambda_0/\lambda)^{\beta}$.
Physically, this represents the cold dust reservoir which contains most of the mass of dust in galaxies' ISM.
The transition between the two piecewise regimes occurs at the wavelength where the derivative of the modified blackbody in log-log space equals $\alpha$.
This piecewise form follows the methodology of e.g. \citet{Blain03a}, \citet{Younger09c}, and \citet{Roseboom13v}. 
The model has the advantage of accounting for potential MIR excesses while also minimizing the number of free parameters, which is very important given the dearth of observations at wavelengths between 8--1000\,\um.
While there are galaxies for which such a model may be incomplete --- for example, dusty, luminous quasars --- the vast majority of galaxies with non-dominant AGN can be modeled using this simple technique.

Previous work by \citet{Casey12a} has shown that SED fits with a modified blackbody plus mid infrared power law produce statistically better fits than fits with IR templates, despite having fewer free parameters. If one's goal is to fit the IR luminosity and peak wavelength, as is our goal in the present work, it is better to use the model we adopt here than to fit templates. Such a model has been used widely in the literature \citep[e.g.][]{Blain02a, Blain03a, Younger09c, Casey12a} because it produces good fits to data and has fewer free parameters than methods like template fitting \citep[e.g.][]{Chary01a, Dale02a, Siebenmorgen07a} and models with power law distributions of dust temperatures \citep[e.g.][]{Kovacs10r}.

The fit parameters that may be fixed when fitting are the power law slope, $\alpha$, the dust emissivity index, $\beta$, and the wavelength where the opacity equals unity, $\lambda_0$. The peak wavelength and IR luminosity are always free parameters.
Consistent with previous works \citep[e.g.][]{Sajina06a, Hayward12a, Spilker16a}, we find that the parameters $\beta$, \td, $\lambda_0$ have significant covariance with one another under the current limitations of long wavelength datasets. Note, however, the covariance does not extend to \lp, which is measured after the fact from the resulting trial fit.

\subsection{Choice of Parameter Priors}
We choose to fix $\lambda_0 = 200$\,\um\ for all galaxies analyzed in this paper for a few reasons. First, our observations -- like the vast majority of all galaxies -- do not have sufficient wavelength coverage near the peak of dust emission to constrain $\lambda_0$ as a free parameter.
Second, we do not have resolved dust continuum measurements for these galaxies.
Third, fixing $\lambda_0 = 200$\,\um\ allows our peak wavelength constraints to be more easily compared with other works in the literature that do the same.
Last, our fitting function is, at face value, phenomenological rather than physical. In reality each galaxy has a distribution of dust temperatures rather than a single `cold' temperature representative of the bulk luminosity of the ISM.
A galaxy with a distribution of luminosity-weighted dust temperatures may emit radiation in such a way that mimics opacity effects from the self absorption factor, $(1-e^{-\tau})$ or a shallower measured dust emissivity index, $\beta$.
Typical values of $\lambda_0$ measured for similarly IR-luminous galaxies or adopted in the literature range from $\approx$100--200\,\um\ \citep[e.g.][]{Blain03a, Draine06a, Conley11a, Rangwala11d, Greve12a, Conroy13a, Spilker16a, Simpson17a, Zavala18a, Casey19a}.

MCIRSED employs Markov chain Monte Carlo (MCMC) to construct posterior distributions for our fit parameters and derive confidence intervals.
We use the Python package, PyMC3 \citep{Salvatier16a}, with its Hamiltonian Monte Carlo fitter (HMC) and the No-U-Turn Sampler, NUTS \citep{Hoffman11a} to efficiently explore parameter space without requiring the user to tune the HMC parameters.
The NUTS algorithm takes advantage of the gradients of the Bayesian likelihoods in order to converge more quickly than other sampling methods for models with high dimensionality.
One of the key strengths of using the PyMC3 package with NUTS is that no prior knowledge or testing of step size or number of steps per sample is required for a given galaxy.
We refer the reader to \citet{Hoffman11a} for the NUTS algorithm and more detail on the tuning phase with the NUTS sampler.

PyMC3 requires the user to specify the model to fit to the data, the prior distibutions for each parameter associated with the model, the number of samples to run, and the number of tuning steps to use.
Tuning is a phase of fitting done prior to sampling in which the step size and number of steps per sample are automatically varied to minimize the sampling time while trying to reach the target acceptance rate.
We choose 3000 tuning steps for all of our fits and exclude those sampling steps from our final analysis.
We find that adding more tuning steps does not change the fits for our galaxies.
We sample the posterior distributions of each galaxy 5000 times because with this number we find our sampling always converges with $>$1000 effective samples.

For the \iras\ sample we use flat bounded priors on $\alpha$, $\beta$, and \td. The bounds are chosen to restrict sampling to parameter space that corresponds to conditions that are physical though not restrictive. 
The dust temperature may not be lower than the temperature of the CMB at the redshift of the galaxy being fitted and not hotter than 1200\,K, a temperature that corresponds to some of the hottest quasars' host galaxies ISMs \citep[e.g.][]{Glikman06k}.

The MIR power law slope, $\alpha$, is constrained to be between $0 < \alpha < 15$. Higher $\alpha$ values asymptotically approach the case of a pure Planck function with $\alpha=15$ being nearly indistinguishable. See the left panel of \autoref{fig:different_alphas} for a plot of example SEDs with a range of $\alpha$ values from 0.5 to 15. The $\alpha=15$ curve and the Planck function are indistinguishable.

The dust emissivity index, $\beta$, is constrained to be between $0.5 < \beta < 5.5$. Some of the highest observed dust emissivity values are $\beta = 3.7$ \citep[][observed in the giant molecular cloud, Sagittarius B2]{Kuan96h}, so we choose a limit beyond that of 5.5. MCIRSED allows the user to set the prior boundaries if one wishes to use higher or lower bounds. For the galaxies analyzed in the present work we find none of the bounds affect our sampling of the posteriors, as would be evidenced by truncated posterior distributions. See the right panel of \autoref{fig:different_alphas} for a plot of example SEDs with $0.5 < \beta < 5.5$.

For the \hatl\ and COSMOS samples, which do not have as high quality constraints as the \iras\ sample, we use the measured average \iras\ sample parameters as a basis, adopting normal priors on $\alpha$ and $\beta$ with medians and standard deviations from the posterior distributions from the \iras\ sample. The adopted values are $\mu_{\alpha} = 2.3$, $\sigma_{\alpha} = 0.5$, $\mu_{\beta} = 1.96$, and $\sigma_{\beta} = 0.43$.
The median emissivity, $\beta = 1.96$ is fully consistent with several recent publications that find similarly steep values for the emissivity spectral index in various high-redshift galaxies with high quality far-IR constraints \citep[e.g.][]{Kato18j, Casey21a}.
We choose to adopt these priors because the minority of galaxies which produce unphysical fits when run with flat priors have more physical fits using normal priors.
For instance, we find that 7.5\% of COSMOS galaxies have $\beta$ that are higher than the highest measured value in the literature of 3.7 \citep{Kuan96h} when fitted with flat bounded priors, however the adoption of normal priors results in much lower $\beta$ values (median 2.03). The reason for the high $\beta$ values is that the SNR of observations that constrain $\beta$ are low. In the absence of firm constraints the Gaussian priors return more reasonable values of $\beta$, rather than something closer to an average of all allowed values between 0.05 and 5.5. For sources with high SNR observations on the RJ tail, the best fit values of $\beta$ are unaffected by the choice of a flat bounded prior or that of a Gaussian prior taken from the posterior of the \iras\ sample fits. Fitting galaxies with Gaussian priors in the case of low SNR observations on the Rayleigh-Jeans tail, for instance, is less restrictive than fixing $\beta$.

\subsection{Correction for Cosmic Microwave Background Emission}
Though it is not necessary to account for CMB heating of dust SEDs in this paper (as all sources are at $z < 2$) we have build MCIRSED to be used for galaxies at any redshift, including those at $z > 5$ where CMB heating has a non-negligible effect on the SED.
MCIRSED allows users to choose whether to correct for the effects of observing a galaxy against a non-negligible CMB temperature. By default the CMB correction is off, but one may toggle it on when calling the MCMC fitting function.
Work by \citet{da-Cunha13a} demonstrates that if the CMB at the redshift of the observed galaxy is non-negligible compared with the dust heating due to the interstellar radiation field, corrections need to be made (1) to the observed dust temperature in order to isolate the dust heating due only to the interstellar radiation field, and (2) to the luminosity of the galaxy observed against the non-negligible background continuum.
When this setting is enabled in MCIRSED, in addition to the parameters mentioned above, the $T_{dust}^{z=0}$ parameter, the temperature the dust would have if bathed in the $z=0$ CMB, and $S_{\nu}^{\rm intr}$, the flux density intrinsic to the source without contribution from the CMB will also be sampled.
These quantities are defined as:

\begin{gather}
    S_{\nu}^{\rm intr} = \frac{S_{\nu}^{\rm obs}}{1-\frac{B_{\nu}(T_{\rm CMB}(z))}{B_{\nu}(T)}},
\end{gather}
where $S_{\nu}^{obs}$ is the observed flux density, $B_{\nu}$ is the Planck function, $T_{\rm CMB}(z)$ is the temperature of the CMB at the redshift of the galaxy, and $T$ is the dust temperature measured from the best fit to the data, as above.

\begin{gather}
    T_{dust}^{z=0} = \left[(T)^{4+\beta} - (T_{\rm CMB}^{z=0})^{4+\beta}[(1+z)^{4+\beta}-1]\right]^{\frac{1}{4+\beta}},
\end{gather}
where $T_{\rm CMB}^{z=0}$ is the temperature of the CMB at $z=0$, and $\beta$ is the dust emissivity index. See \citet{da-Cunha13a} for more detail.

\bibliography{p2}
% \bibliographystyle{aasjournal}



\end{document}